\documentstyle[graphicx,floats,prl,aps,rotate]{revtex}
\begin{document}

\newcommand{\nucleus} [2] {\mbox{${}^{#2}{{\rm #1}}$}}

\draft
\title{
Neutrino-nucleus reactions and nuclear structure\footnote{to be published 
in J. Phys. G. {\bf 29}, 1 (2003).}
}
\author{E. Kolbe$^1$, K. Langanke$^2$, G. Mart\'{\i}nez-Pinedo$^3$, 
and P. Vogel$^4$}

\address{$^1$ Department f\"ur Physik und Astronomie der Universit\"at
Basel, Basel, Switzerland \\,
  $^2$Institut for Fysik og Astronomi, {\AA}rhus Universitet, DK-8000
  {\AA}rhus C, Denmark\\
  $^3$  Institut d'Estudis Espacials de Catalunya, Edifici Nexus,
Gran Capit\`a 2, E-08034 Barcelona, Spain \\
and Instituci\'o Catalana de Recerca i Estudis Avan\c{c}ats,
Llu\'{\i}s Companys 23, E-08010 Barcelona, Spain \\
  $^4$ Department of Physics, California Institute of Technology,
Pasadena, CA 91125, USA 
}

\date{\today}
\maketitle

\begin{abstract}
The methods used in the evaluation of the neutrino-nucleus cross section
are reviewed. Results  are shown
for a variety of targets of practical importance. Many of the described
reactions are accessible in future experiments with neutrino sources
from the pion and muon decays at rest, which might be available at
the neutron spallation facilities. Detailed comparison between the 
experimental and theoretical results would establish benchmarks
needed for verification and/or parameter adjustment of the nuclear
models. Having a reliable tool for such calculation is of great
importance in a variety of applications, e.g. the neutrino oscillation
studies, detection of supernova neutrinos, description of the neutrino
transport in supernovae, and description of the r-process nucleosynthesis. 
\end{abstract}

\section{Introduction}

There is now a convincing evidence that neutrinos are massive and the
existence of neutrino oscillations have been convincingly demonstrated.
This conclusion is based to a large extent on the observation of the
neutrino induced reactions on complex nuclei, which also
play essential roles in various aspects of nucleosynthesis as well as
supernova collapse and supernova neutrino detection. Most of them
have not been studied experimentally so far and their
cross sections, which are needed in all applications,
are typically based on nuclear theory. Spallation
neutrino sources with their significant neutrino fluxes
represent a unique opportunity to establish several benchmark
measurements of the most significant neutrino-nucleus reactions.
These measurements, in turn, can be used to gauge the accuracy
and reliability of the corresponding nuclear models.

In this work we review selected theoretical results
of particular importance for neutrino detection, supernovae,
and nucleosythesis.
Theoretical description of the neutrino induced
reactions is a challenging proposition, since the energy scales of interest
span a vast region, from the few MeV for solar neutrinos, to tens 
of MeV for the interpretation of experiments with the muon and pion
decay at rest and the detection of supernova neutrinos, 
to hundreds of MeV or few  GeV for the detection
atmospheric neutrinos.  While reactions induced by
low-energy neutrinos are sensitive to details of nuclear structure,
GeV neutrinos, like other weak probes of similar energy, interact 
dominantly with individual nucleons in the nucleus, 
which can then be treated  as an ensemble of
non-interacting but bound protons and neutrons.

First, let us briefly review the general formalism adopted for
the analysis of the charged current reactions,
\begin{equation}
\nu_e + _{Z} \! X_{N} \rightarrow  _{Z+1}X^{*}_{N-1} + e^{-}
\end{equation}
and its analogs with $\bar{\nu}_e$ as well as with muon neutrinos
and antineutrinos. The formalism can be easily modified for the
neutral current reactions
\begin{equation}
\nu + _{Z} \! X_{N} \rightarrow  _{Z}X^{*}_{N} + \nu'
\end{equation}

In the derivation of the relevant cross sections we follow the prescription
given by Walecka \cite{Walecka} which is based on  
the standard current-current form for
the weak interaction Hamiltonian governing
these reactions. After a multipole expansion of the weak nuclear current and
application of the extreme relativistic limit 
(final lepton energy
$E_{\ell} >\!>$ lepton mass $m_{\ell}c^{2}$)
the neutrino (antineutrino) cross
section for excitation of a discrete target state is given by
\cite{Walecka,Co72}:
\begin{equation}
   \label{Cr01a}
\left ( \frac{d \sigma_{i \rightarrow f}}
{d \Omega_{\ell}} \right )_{\nu, ~\bar{\nu}} = 
\frac{(G_F V_{ud})^2 \cdot p_{\ell} E_{\ell}}{ \pi} \cdot
\frac{\cos^{2} \frac{\Theta}{2}}{(2 J_{i} + 1)} \cdot F(Z\pm1,
\epsilon_{\ell}) \cdot \left [ \sum_{J=0}^{\infty} \sigma_{CL}^J +
                    \sum_{J=1}^{\infty} \sigma_{T}^J \right ]
\end{equation}
where
\begin{equation} 
   \label{Cr01b}
\sigma_{CL}^J =   | < J_{f} |\!| \tilde{M}_{J} (q)
+ \frac{\omega}{{q}} \tilde{L}_{J}(q) |\!| J_{i} >|^{2} 
\end{equation}
and 
\begin{equation}
   \label{Cr01ic}
\begin{array}{ll}
\sigma_T^J =
\mbox{} & \left ( -\frac{q^{2}_{\mu}}{2 {q}^{2}} + \tan^{2}
\frac{\Theta}{2} \right ) \times \left [
 |< J_{f} |\!|
\tilde{J}_{J}^{mag}(q) |\!| J_{i} >|^{2} + |<J_{f}|\!| \tilde{J}^{el}_{J}
(q) |\!| J_{i} > |^{2} \right ] \\
\mbox{} &  \mp \tan \frac{\Theta}{2} \sqrt{\frac{-q^{2}_{\mu}}{
{q}^{2}}
+ \tan^{2} \frac{\Theta}{2}} \times \left [
2 Re < J_{f} |\!| \tilde{J}^{mag}_{J}(q) |\!| J_{i} > <J_{f} |\!| \tilde{J}
^{el}_{J}(q) |\!| J_{i}>^{*} \right ]  \, .
\end{array}
\end{equation}
Here $\Theta$ is the angle
between
the incoming and outgoing lepton, and $q_{\mu} = (\omega, \vec{q})$ $(q =
|\vec{q}|)$ is the four-momentum transfer. The minus-(plus) sign in Eq. (5)
refers to the
neutrino (antineutrino) cross section. The quantities
$\tilde{M}_{J}, \tilde{L}_{J},
\tilde{J}^{el}_{J}$ and
$\tilde{J}^{mag}_{J}$ denote the multipole operators for the charge, 
the longitudinal and the transverse electric and magnetic parts of the
four-current, respectively. Following Refs. \cite{Walecka,Don-Hax}
they can be written in terms of one body operators in the nuclear
many-body Hilbert space. The cross
section involves the reduced matrix elements of these operators
between the
initial state $J_{i}$ and the final state $J_{f}$.
(See Refs. \cite{DP79,Kolbe96} for the slightly more
complicated formula valid also for
nonrelativistic final lepton energy.) 

For low energy electrons and positrons the Fermi
function $F(Z, E_{\ell})$ accounts for the Coulomb interaction between
the final charged lepton and the residual nucleus in the charged-current 
processes.
We use the Coulomb correction derived by numerical solution
of the Dirac equation for an extended nuclear charge \cite{Be82}:

\begin{equation}
   \label{Cr02}
F(Z, E_{\ell})  = F_{0}(Z, E_{\ell}) \cdot L_{0} ~,
{\rm {~~with~~}}  F_{0}(Z, E_{\ell}) = 4(2p_{l}R)^{2(\gamma - 1)}
\left | \frac{\Gamma (\gamma + i y)}{\Gamma(2 \gamma + 1)} \right |^{2}
\cdot e^{\pi \cdot y} \, .
\end{equation}
Here $Z$ denotes the atomic number of the residual nucleus in the final
channel, $E_{\ell}$
the total lepton energy (in units of $m_{\ell}c^{2}$) and $p_{l}$ the lepton
momentum (in units of $m_{\ell}c$),
$R$ is the nuclear radius (in units of $\frac{\hbar}{m_{\ell}c}$)
and $\gamma$ and $y$ are given by ($\alpha =$ fine structure constant):
\begin{equation}
\gamma = \sqrt{1 - (\alpha \cdot Z)^{2}} \qquad ~,~
{\rm and} \qquad y = \alpha \cdot Z \cdot 
\frac{E_{\ell}}{p_{\ell}} ~.
\end{equation}
The numerical factor $L_{0}$ in (6),
which describes the finite charge
distribution and screening corrections, is nearly constant($\approx 1.0$),
and can be well approximated by a weakly decreasing
linear function in $p_{\ell}$.

At higher energies, and for muons at essentially all energies,
the Fermi function valid for $s$-wave leptons
is a poor approximation for the
Coulomb effect since higher partial waves also contribute
for $p_{\ell}R \ge 1$. Guided by the distorted-wave
approximation of quasielastic electron scattering,
we treat in that case the Coulomb effects in the `Effective Momentum
Approximation' in which the outgoing lepton momentum
$p_{\ell}$ is replaced by the effective momentum
\begin{equation}
p_{eff} = \sqrt{E_{eff}^2 - m_{\ell}^2}, ~~E_{eff} = E_{\ell} - V_C(0) ~,
\end{equation}
where $V_C(0) = 3 e^2 Z / 2 R$ is the Coulomb potential at the origin. 
In the work presented here,
the Coulomb effect is taken into account not only by using the
effective momentum, but also by replacing the phase space factor
$p_{\ell} E_{\ell}$ by $p_{eff} E_{eff}$ (see also \cite{Engel98}
where this procedure is called modified effective momentum approximation,
and shown to work quite well).
In practice we use a smooth interpolation between these two 
regimes of treatment of the Coulomb effects.

We calculate the differential cross section (3) as a function
of the initial neutrino energy $\epsilon_{\nu}$, 
the excitation energy of the nucleus
$\omega$ and the scattering angle $\Theta$. 
The three-momentum transfer $q \equiv |\vec{q}|$ is equal to 
\begin{equation}
{q} = \sqrt{(E_{\nu} - p_{\ell})^2 + 4 E_{\nu} \cdot p_{\ell} 
\sin^{2} \frac{\Theta}{2}} ~\simeq~ 
\sqrt{\omega^{2} + 4 E_{\nu} \cdot (E_{\nu} - \omega)
\sin^{2} \frac{\Theta}{2}} \, ,
\end{equation}
where the last expression is valid in the relativistic limit 
($E_{\ell} \gg m_{\ell}$) for the final lepton.
The total cross section is obtained from the differential cross sections
by summing (or integrating) over all possible final nuclear
states and by numerical integration  over the angles.

As explicitly used in the derivation of the cross section formula above,
in neutrino-induced reactions the nucleus is excited by multipole
operators $O_\lambda$ which scale like $(qR/\hbar c)^\lambda$, 
where $R$ is the nuclear radius ($R \sim 1.2 A^{1/3}$ fm). 
As the momentum transfer
is of the order of the neutrino energy $E_{\nu}$,
neutrino-nucleus reactions involve  multipole
operators with successively higher rank $\lambda$ with increasing
neutrino energy. Since the nuclear Hamiltonian does not commute with
$O_\lambda$, the response of the operator is fragmented over many
nuclear states. 
However, for each multipole
most of the strength resides in a collective excitation,
the giant resonance, with a width of a few MeV. The centroids of
the giant resonances of the various multipoles  grow in 
energy with increasing rank, roughly like $\lambda \hbar \omega$, 
where $\hbar \omega \simeq 41/A^{1/3}$ MeV 
is a typical energy shell splitting
in the nucleus. Furthermore, since the phase space is proportional to 
$p_{\ell} E_{\ell}$ the higher outgoing lepton energies
are preferred.
This suggests that the average nuclear excitation energy 
$\bar{\omega}$ lags behind the increasing
neutrino energy, i.e., for sufficently 
large neutrino energies the inequality
$\bar{\omega} < E_{\nu} \simeq E_{\ell}$ holds. 
As a consequence, for neutrino
energies significantly larger than the energies of the corresponding
giant resonances, 
the neutrino-induced cross sections will depend on the total
strength of the multipole excitation and its centroid energy, but 
will be less sensitive to
its detailed energy distribution. Finally, at very high neutrino
energies the neutrino will see the nucleus as an ensemble of independent
bound nucleons and will interact with individual nucleons. In that regime
we will consider only the quasielastic channel in which the
struck nucleon is ejected. This is the channel
most widely used in neutrino detectors.

From these general considerations we can identify three different energy
ranges with quite different demands on the details with which
the nuclear structure should be treated:

i) For relatively low neutrino energies, 
comparable with the nuclear excitation
energy, neutrino-nucleus reactions are very sensitive to the appropriate
description of the nuclear response. Thus, low energy neutrino
scattering requires a nuclear model which reproduces the important
correlations among nucleons. The model of choice is the nuclear shell
model, which accounts for nucleon-nucleon correlations via an effective
interaction within a fixed model space for the valence nucleons.
Nowadays, complete diagonalization for the lowest states in medium-mass
nuclei (up to $A \sim 60$) is achievable in complete $0 \hbar \omega$
model spaces, i.e., considering all configurations of the valence
nucleons in a full harmonic oscillator shell. For lighter nuclei, like
$^{16}$O, complete diagonalization can be performed in larger model
spaces ($4 \hbar \omega$ for $^{16}$O, $6 \hbar \omega$ for $^{12}$C).
Importantly, the shell model calculations have been proven to indeed
reproduce the allowed (Fermi, Gamow-Teller) response for those nuclei
for which diagonalizations can be performed in sufficiently large model
spaces. For the lighter nuclei, where multi-shell
calculations can be performed, the shell model also nicely describes
forbidden transitions. An overview of recent shell model developments
and applications is given in \cite{Caurier04}.

ii) The Random Phase Approximation (RPA) 
has been developed to describe the collective excitation of
a nucleus by considering the one-particle one-hole excitations of the
correlated ground state. In the standard RPA, all excited states are
treated as bound states, leading to a discrete excitation spectrum. In
the Continuum RPA (CRPA)(see e.g. \cite{CRPA,Kolbe92} and references therein)
the final states have the appropriate scattering
asymptotics for energies above the nucleon-emission thresholds;
consequently the excitation spectrum in the CRPA is continuous. The RPA
or CRPA are the methods of choice at intermediate neutrino 
energies where the
neutrino reactions are sensitive 
dominantly to the total strength and the energy
centroids of the giant resonances.

iii) At high incoming energies neutrinos, like other electroweak probes,
scatter `quasi-freely' on individual nucleons. 
The remaining nucleons can be treated
as (non-interacting) spectators. This situation is realized in the Fermi
gas model \cite{Moniz}
where a full relativistic treatment of the
hadronic weak vertex is included. The Fermi motion and the
binding energy are characterized by  parameters
that can be determined from electron scattering experiments.
In this approximation the expansion in multipoles is no longer
necessary. The nuclear form factors for the quasi-free scattering
are expressed in terms of the single nucleon matrix elements
which  depend only on the four-momentum transfer $q^2$ and on the
nuclear momentum distribution.  

Above we pointed out the important role of collective excitations.
The centroid position of these excitations deviates noticeably
from the independent particle estimate of $\lambda \hbar \omega$
due to the residual particle-hole interaction. Thus, empirical
evidence or nuclear structure calculation is needed to determine
their energy. On the other hand, the total strength is often
fixed by sum rules in an essentially model independent way. 
Well known examples of such sum rules is the Ikeda sum rule for the
Gamow-Teller strength
\begin{equation}
\sum_i B(GT;Z \rightarrow Z+1)_i - \sum_i B(GT;Z \rightarrow Z-1)_i
= 3 (N- Z) ~,
\end{equation}
or the Thomas-Reiche-Kuhn sum rule for the dipole strength
\begin{equation}
\sum_i(E_i - E_0)B(E1;0 \rightarrow i) = \frac{9}{4 \pi} 
\frac{\hbar^2}{2 M_p} \frac{N Z}{A} e^2 ~.
\end{equation} 
Even though both of these sum rules could be violated to some extent
(e.g., when the internal structure of the nucleons is not properly treated), 
the dependence
on the neutron and proton numbers $N$ and  $Z$ remains valid.
Note that the Ikeda sum rule involves the difference of the strengths.
However, in nuclei with neutron excess, $ N > Z$, the second term,
the total strength $B(GT;Z \rightarrow Z-1)$ is much smaller than
the first one, and so the sum rule determines the strength in the 
$(p,n)$ channel. In symmetric nuclei with $N = Z$ the Ikeda sum rule,
unfortunately, does not help in fixing the Gamow-Teller strength.

To calculate the various partial neutrino-induced reaction cross
sections for neutrino-induced reactions  we assume a
two-step process. In the first step we calculate the charged current
($\nu_l,l^-$) and (${\bar \nu_l},l^+$)
cross sections (where $l=e$ or $\mu$), 
or the neutral current cross section ($\nu,\nu'$)
as a function of excitation energy in the final nucleus. These
calculations are performed within the 
RPA or CRPA and considering all multipole operators up to a certain
$J$ and both parities. 
In the second step one calculates for each
final state with well-defined energy the
branching ratios into the various decay channels using the statistical
model code SMOKER \cite{Friedel}. As possible final states in the
residual nucleus the SMOKER code considers the experimentally known
levels supplemented at higher energies by an appropriate level density
formula \cite{Friedel}. Proton,
neutron, $\alpha$ and $\gamma$ emission are included in the code
as decay channels. 
If the decay leads to an
excited level of the residual nucleus, 
the branching ratios for the decay of this state is calculated in an
analogous fashion \cite{Langanke96}. 
Keeping track of the energies of the ejected
particles and photons during the cascade, and weighting them with
appropriate branching ratios and the corresponding primary 
charged- or neutral-current cross sections, we determine the various
partial particle emission cross sections. 

\section{Comparison of different methods}
Here we demonstrate, using the neutrino interaction with $^{16}$O
as an illustration, how different theoretical methods can be used
at different neutrino energies. We show that at certain transition
energy intervals the corresponding methods give essentially identical
results. 

\subsection{Shell model versus CRPA}

\begin{figure}[ht]
  \begin{center}
    \leavevmode
    \includegraphics[height=10.5cm]{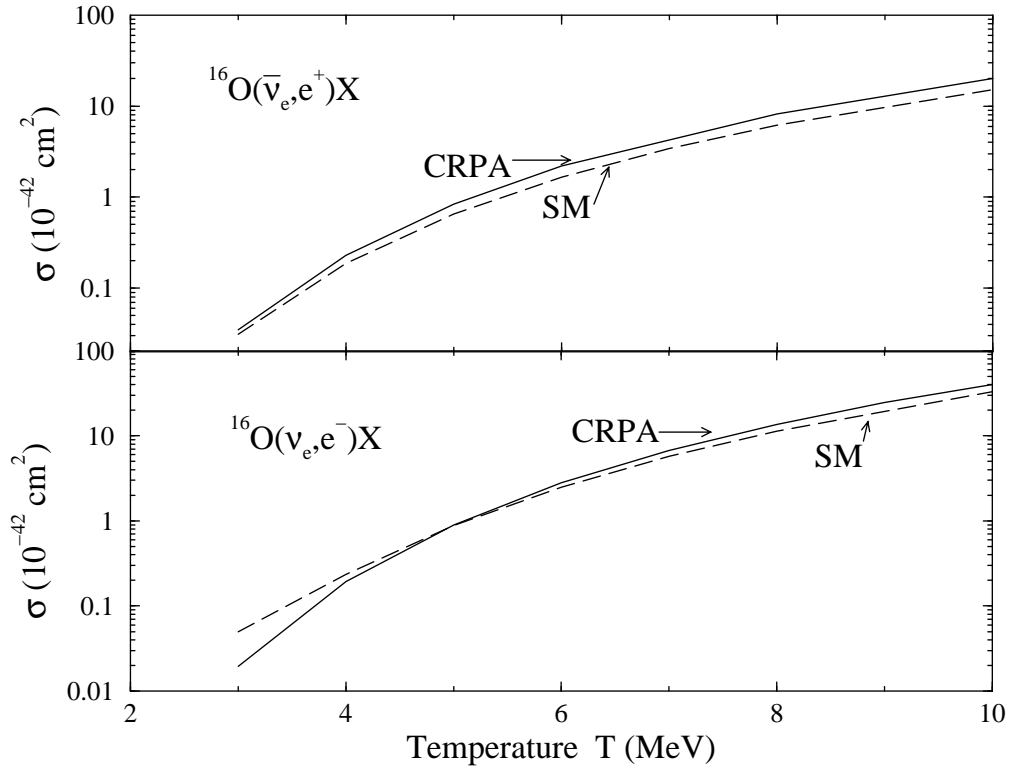}
    \caption{ Comparison of the CRPA (full lines)
and shell model (dashed lines) cross sections.
The upper panel is for the $\bar{\nu}_e$
induced reaction and the lower one is for the reaction induced by $\nu_e$.
        }
    \label{fig:SM-CRPA_cr}
  \end{center}
\end{figure}

We first consider the description of the charged current reactions
on $^{16}$O at relatively low energies. The reaction thresholds are
15.4 MeV for the $^{16}$O$(\nu_e, e^-)^{16}$F (which is unbound) 
and 11.4 MeV for the $^{16}$O$(\bar{\nu}_e, e^+)^{16}$N reaction. 

The shell model evaluation
of the cross sections was performed back in 1987 by Haxton \cite{Haxton87}.
In that work the low lying positive parity states were described in a full
2 $\hbar \omega$ shell model. The transitions to negative parity states
were described using the effective density matrices, scaled to describe
measured form factors from electron scattering. 

The shell model results can be compared to the CRPA. The CRPA calculations used
the finite range residual force based on the Bonn potential, and all
multipole operators with $J \le 9$ and both parities were included.
The free nucleon form factors were used, with no quenching. The procedure
was tested by evaluating the total muon capture rates 
(dominated by the negative
parity multipoles) for $^{12}$C, $^{16}$O and $^{40}$Ca \cite{Kolbe94},
as well as the partial capture rates to the bound $0^-, 1^-$ and the $2^-$ 
ground state in $^{16}$N \cite{Kolbe94,footnote}. Good agreement with these
muon capture rates tests the method at momentum transfer 
$q \sim m_{\mu} \sim 100$ MeV.

   We compare the cross sections evaluated by the two methods in Fig.
\ref{fig:SM-CRPA_cr}, where we show the cross sections evaluated in
both methods and averaged over the Fermi-Dirac distribution
corresponding to the temperature $T$ and vanishing chemical
potential. 
The agreement is excellent suggesting
that both methods are 
capable of describing the weak reaction rates in this energy regime,
provided that they can be successfully tested on relevant
quantities, such as the muon capture rates, nuclear photoabsorption
cross section, or inelastic electron scattering leading to the
states populated by the weak processes.

\begin{figure}[htb]
  \begin{center}
    \leavevmode
    \includegraphics[height=10.5cm]{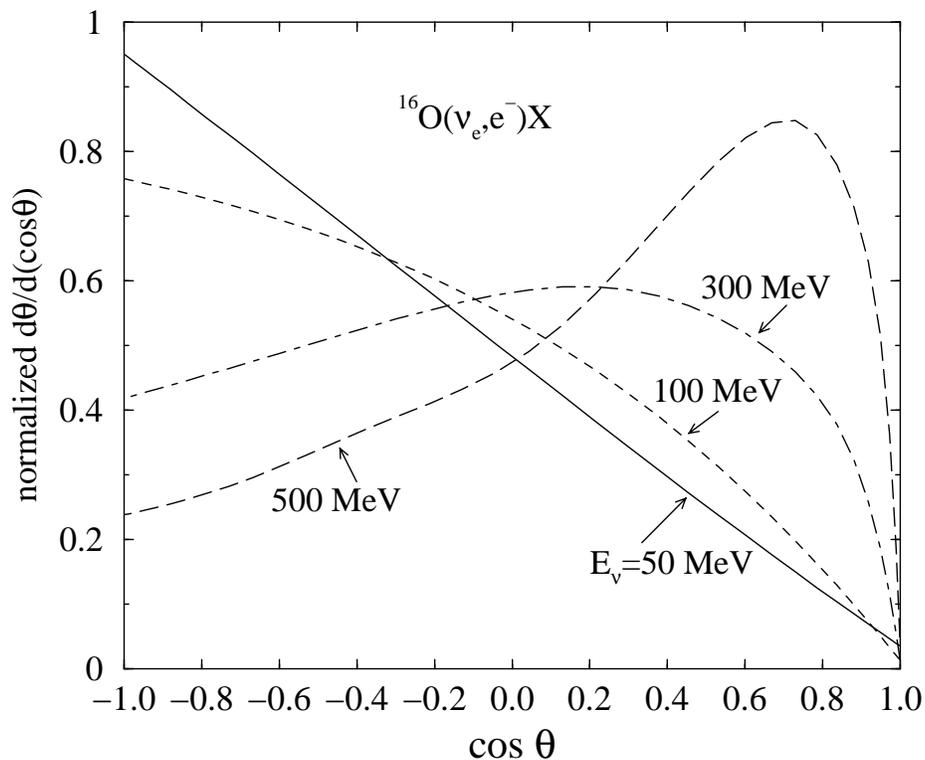}
    \caption{ The CRPA angular distributions.}
    \label{fig:CRPA_ang}
  \end{center}
\end{figure}

The angular distribution of the emitted electrons with respect to
the incoming neutrino beam is shown in Fig. \ref{fig:CRPA_ang}.
Note the electron emission  is predominantly
in the backward direction at low energies (also 
obtained in the nuclear shell model), but it gradually
changes to the forward one at higher energies. Thus, for 
$E_{\nu} \ge$ 500 MeV the direction of the electron can be
used to determine the direction of the incoming neutrino.

We have thus identified the energy region, somewhere near about 50 MeV
of neutrino energy, where the two discussed methods, the nuclear shell
model and the CRPA, give essentially identical results. 
For lower energies the nuclear shell model is the method of
choice. As the 
energy increases, the shell model calculations become increasingly difficult.
The number of states increases rapidly, and the effective interaction
to be used becomes more uncertain. However, as we argued above, at higher
neutrino energies, above, say, $E_{\nu} \ge$ 100 MeV, the details of the
nuclear correlations become less important and what matters are the
positions and strengths of giant resonances. The CRPA is capable of
describing these quantities and thus, in our opinion, it is the
method of choice at the intermediate neutrino energy range, 
approximately $100 {\rm ~MeV} \le E_{\nu} \le 500 {\rm ~MeV}$.
As the energy increases further, the CRPA calculations become 
computationally more difficult (more multipoles and higher
nuclear excitation energies must be included). At the same time,
the nuclear response, at least for the quasielastic regime, 
becomes simpler.

\subsection{ CRPA versus Relativistic Fermi Gas model}

\begin{figure}[htb]
  \begin{center}
    \leavevmode
    \includegraphics[width=0.8\columnwidth,height=12.0cm]{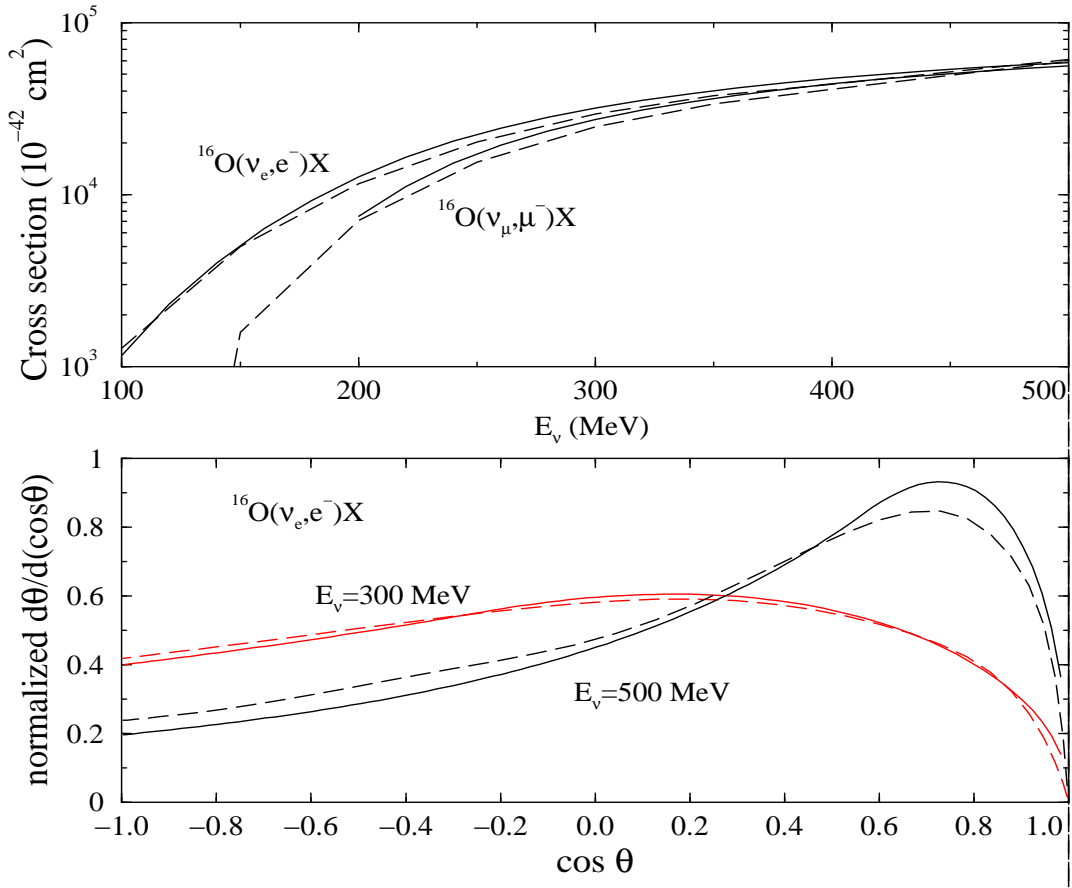}
    \caption{ Comparison of  the relativistic Fermi gas model (full lines)
 and the  CRPA (dashed lines). The parameters of the Fermi gas model were
$p_f = 225$ MeV and $e_b = 27$ MeV. The cross sections are shown in the
upper panel for the two indicated reactions. The angular distributions for
$E_{\nu}$ = 300 and 500 MeV are shown in the lower panel.
The CRPA results are shown with full lines, RFG with dashed lines.  }
    \label{fig:fg_crpa_cr}
  \end{center}
\end{figure}

The CRPA and Relativistic Fermi Gas model (RFG) 
methods are compared in Fig. \ref{fig:fg_crpa_cr} and agree
remarkably well in both the total quasielastic cross section,
as well as in the angular distribution of the outgoing electrons. The
latter is particularly important, because the zenith angle distribution
of the atmospheric neutrinos is based on the assumption that one can
deduce the incoming neutrino direction (and hence its flightpath)
from the direction of the observed charged lepton. Our CRPA
calculations confirm that, indeed, below about 500 MeV of neutrino
energy the emitted electron (or muon) are essentially uncorrelated
with the direction of the incoming neutrino. Above these energies,
the emitted lepton moves dominantly in the direction of the incoming
neutrino, hence one can, statistically, correlate the two. This
tendency, naturally, becomes more pronounced at higher $E_{\nu}$ values.

Let us stress that no attempt was made to adjust the Fermi gas
model parameters to obtain the agreement demonstrated in Fig. 
\ref{fig:fg_crpa_cr}; these are just standard values of the
Fermi momentum $p_f$ and of the parameter $e_b$ which characterizes
the average nucleon binding energy.

Thus, we  conclude that at energies $E_{\nu} \simeq 300$ MeV
one can safely switch from the CRPA to the simpler Fermi gas
model description of the quasielastic charged current reactions.
However, it should be stresses that at low energies there
are important checks in the form of muon capture rates as well
as (for $^{12}$C) neutrino induced reaction for the neutrino
beams from the muon and pion decay at rest. Similar tests do not
exist, or were at least less exploited, for $\sim 1$ GeV neutrino
energies.

\section{Neutrino $^{12}$C interaction}

The nucleus $^{12}$C is particularly important for the
study of the neutrino-nucleus scattering.
Liquid scintillator detectors, for example
KARMEN and LSND, contain hydrogen and $^{12}$C nuclei. 
Therefore, a number
of experimental results exist in this case
as byproducts of the neutrino oscillation searches
performed with these detectors.

The measurements include charged-current reactions induced by both
electron- \cite{Karmen,LSND} and muon-neutrinos \cite{LSND},
exciting both the ground and continuum states in $^{12}$N.  
As discussed below, the inclusive
cross section for $^{12}$C($\nu_e,e$)$^{12}$N$^{*}$ 
with the $\nu_e$ from the muon decay at rest (DAR)
\cite{Karmen,LSND,Krakauer},
agrees well with calculations, while
in contrast, there is a discrepancy between
calculations\cite{Kolbe94,Kolbe95,Oset,Oset98} 
(with some notable exceptions \cite{Mintz,Auerbach}) 
and the measured\cite{LSND} inclusive cross
section for  $^{12}$C($\nu_\mu,\mu$)$^{12}$N$^{*}$,
which uses higher energy neutrinos from pion decay-in-flight (DIF). 
The disagreement is somewhat disturbing
in light of the simplicity of the reaction
and in view of the fact that parameter-free calculations, such as
those in\cite{Kolbe94,Kolbe95,Kolbe99a}, describe well other weak processes
governed by the same weak current nuclear matrix elements.
Moreover, as shown in the following subsection,
the exclusive reactions populating the ground state of the final nucleus,
$^{12}$C($\nu_e,e$)$^{12}$N$_{gs}$ and
$^{12}$C($\nu_\mu,\mu$)$^{12}$N$_{gs}$,
and the neutral current reaction $^{12}$C($\nu_e,\nu_e '$)
$^{12}$C(15.11 MeV) have been measured\cite{Karmen,LSND} as well,
and agree well with theoretical expectations.

\subsection{Exclusive reactions}

Among the states in the final nucleus $^{12}$N,
which is populated by the charged current reactions
with beams of $\nu_e$ or $\nu_{\mu}$,
the ground state $I^{\pi} = 1^+$
plays a special role. It is the only
bound state in $^{12}$N, and can be recognized
by its positron decay ($T_{1/2} = $ 11 ms) back to $^{12}$C.
Moreover, the analog of the $^{12}$N$_{gs}$, the $I^{\pi} = 1^+$ state
with isospin $T = 1$ at 15.11 MeV in $^{12}$C, can be populated by the
neutral current neutrino scattering, and is recognizable by its emission
of the 15.11 MeV photon. Finally, even though there are several bound
states in $^{12}$B,  its ground state, the analog of the other two
($I^{\pi}, T)= (1^+,1$) states, is the state most strongly populated
in muon capture on $^{12}$C. Again, the population of the bound states
in $^{12}$B can be separated from the continuum by observing its electron
decay ($T_{1/2} = $ 20.2 ms).

Theoretical evaluation of the exclusive
cross sections is constrained by the obvious
requirement that the same method, and the same parameters, must also describe
the related processes, i.e. the positron decay of $^{12}$N,
the $\beta$ decay of $^{12}$B, the $M1$ strength of the
15.11 MeV state in $^{12}$C, and the partial muon
capture rate leading to the ground state of $^{12}$B.
It turns out that this
requirement essentially determines the neutrino induced cross section for
the energies of present interest. It does not
matter which method of calculation
is used, as long as the constraints are obeyed.

The comparison between the measured and calculated values is shown in
Table \ref{tab:gs}. There, three rather different methods of calculation
were used, all giving excellent agreement with the data.

\begin{table}[h]
\caption {\protect Comparison of calculated and measured cross sections, in
units of $10^{-42}$cm$^{-2}$ and averaged over
the corresponding neutrino spectra, for the neutrino induced transitions
$^{12}$C$_{gs} \rightarrow ^{12}$N$_{gs}$ and $^{12}$C$_{gs} \rightarrow
^{12}$C(15.11 MeV). For the decay at rest the $\nu_e$ spectrum is normalized
from $E_{\nu}$ = 0, while for the
decay in flight the $\nu_{\mu}$ and $\bar{\nu}_{\mu}$
spectra are normalized from the corresponding threshold.
See the text for explanations.}
\label{tab:gs}
\begin{center}
\begin{tabular}{|lccc|}
  & $^{12}$C($\nu_e,e^-)^{12}$N$_{gs}$  &
$^{12}$C($\nu_{\mu},\mu^-)^{12}$N$_{gs}$  &
 $^{12}$C($\nu,\nu')^{12}$C(15.11)  \\
  &  decay at rest   & decay in flight  & decay at rest \\
\hline
experiment\cite{Karmen} & 9.4$\pm0.5\pm0.8$ &  - &11$\pm$0.85$\pm1.0$ \\
experiment\cite{LSND} & 9.1$\pm0.4\pm0.9$  & 66$\pm10\pm10 $ & - \\
experiment\cite{Krakauer} & 10.5$\pm1.0\pm1.0$ &  - & - \\
Shell model \cite{triad} & 9.1 & 63.5  & 9.8 \\
CRPA \cite{Kolbe94,Kolbe95} & 8.9 & 63.0 & 10.5 \\
EPT \cite{Fukugita} & 9.2 & 59 &  9.9 \\
\end{tabular}
\end{center}
\end{table}

The first approach is a restricted shell-model calculation.
Assuming that all structure in the considered low-lying states
is generated by the valence nucleons in the
$p$-shell, and that the two-body currents (pion-exchange
currents) are negligible, there are only four one-body densities (OBD)
which fully describe all necessary nuclear matrix elements.
In this case, it is necessary to use
the one-body densities chosen (ad hoc) in such a way that all 
the auxiliary data mentioned above 
are correctly reproduced. This then gives the
results listed in line 4 of Table \ref{tab:gs}.

Effects of configurations beyond the $p$ shell might explain the need
for the renormalization of the one-body densities produced by a
reasonable $p$-shell Hamiltonian.  Therefore, the rates of all
the reactions are also evaluated in the Random
Phase Approximation (RPA), which does include multishell correlations, while
treating the configuration mixing within the $p$ shell only crudely.
Again an adjustment is needed (a ``quenching'' of all matrix elements by
an universal, but substantial, factor 0.515).
However, the neutrino cross sections
in line 5 of  Table \ref{tab:gs} agree with the measurements very well.

The third approach is the ``elementary-particle treatment" (EPT).  Instead of
describing nuclei in terms of nucleons, the EPT
considers them elementary and describes transition matrix elements in
terms of {\it nuclear} form factors deduced from experimental data.
The EPT approach was extended in Ref. \cite{triad} to the
higher neutrino energies relevant to the LSND decay-in-flight $\nu_{\mu}$'s
by appropriately including the lepton mass.

\begin{figure}
  \begin{center}
    \leavevmode
    \includegraphics[height=8.5cm]{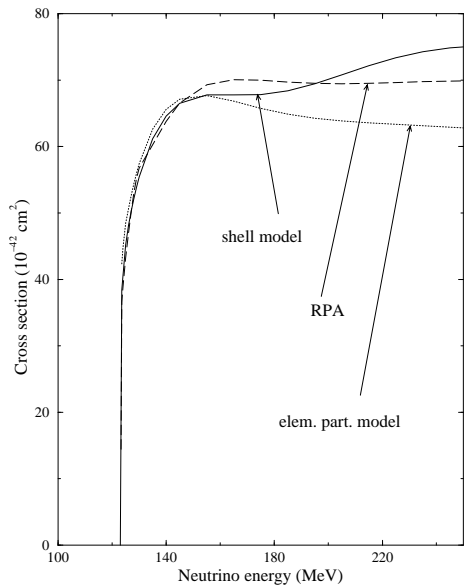}
\caption{Energy dependence of the cross section for the
reaction
$^{12}$C + $\nu_{\mu} \rightarrow ^{12}$N$_{g.s} + \mu^-$.}
\label{fig:gs}
\end{center}
\end{figure}

An example of the energy dependence of the exclusive cross section
is shown in Fig. \ref{fig:gs} for the $\nu_{\mu}$ induced exclusive reaction.
As one can see, the cross section raises sharply from its threshold
($E_{thr} = $ 123 MeV) and soon reaches its saturation value, i.e.,
it becomes almost energy independent. This means that the yield of
the $^{12}$C + $\nu_{\mu}$ reaction essentially measures just the
flux normalization above the reaction threshold. At the same time,
the yield is insensitive to the energy distribution of the muon
neutrinos in the beam.

\subsection{Inclusive reactions}

The inclusive
reactions $^{12}$C($\nu_e,e$)$^{12}$N$^{*}$,
with $\nu_e$ neutrinos from the muon decay-at-rest
and $^{12}$C($\nu_\mu,\mu$)$^{12}$N$^{*}$
with the higher energy $\nu_{\mu}$ neutrinos from the pion
decay-in-flight populate not only the ground
state of $^{12}$N but also the continuum states.
The corresponding cross sections involve folding over
the incoming neutrino spectra and integrating over the
excitation energies in the final nucleus.
By convention, we shall use the term ``inclusive'' for the
cross section populating only the continuum (i.e., without
the exclusive channel) for $^{12}$C($\nu_e,e$)$^{12}$N$^{*}$
with the decay-at-rest $\nu_e$, while for the reaction
$^{12}$C($\nu_\mu,\mu$)$^{12}$N$^{*}$ with the decay-in-flight
$\nu_{\mu}$ the term is used for the total cross section
(the exclusive channel then represents only a small fraction
of the total).

Muon capture, $^{12}$C($\mu, \nu_{\mu}$)$^{12}$B$^{*}$,
belongs also to this category. It involves momentum transfer
of $q \approx m_{\mu}$, intermediate between the two neutrino
capture reactions above. Since $^{12}$B and $^{12}$N are
mirror nuclei, all three reactions should be considered
together. In this case again the term ``inclusive'' will be used
only for the part of the rate populating the continuum in $^{12}$B.

\begin{figure}
  \begin{center}
    \leavevmode
    \includegraphics[height=8.5cm,angle=90]{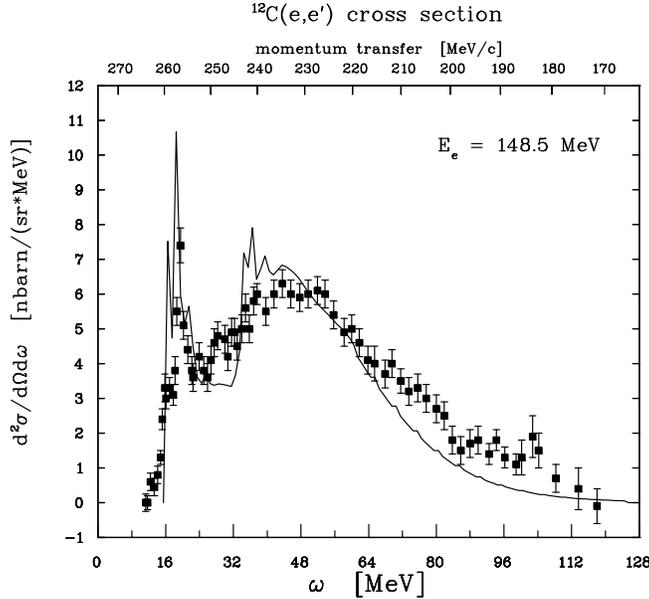}
\caption{ Data (points with error bars) and calculated cross section
for the inclusive electron scattering on $^{12}$C as a function of the
excitation energy $\omega$. The corresponding momentum transfer
is displayed on the upper scale.}
\label{fig:elsc}
  \end{center}
\end{figure}

Which theoretical approach should one use in order to describe
such reactions?  One possibility is to use the
continuum random phase approximation (CRPA). The method has been
used successfully in the evaluation of the nuclear response to
weak and electromagnetic probes \cite{CRPA}. In particular, it was
tested, with good agreement, in the calculation of the inelastic electron
scattering\cite{Leiss} on  $^{12}$C involving very similar excitation energies
and momentum transfers as the weak processes of interest.
As an example  Fig. \ref{fig:elsc} shows  the comparison of the
experimental data and the results of the CRPA for the inclusive
electron scattering\cite{Kolbe96p}. One can see that the CRPA describes quite
well both the magnitude and shape of this cross section over the
entire range of excitation energies and momentum transfers.

For muon capture the CRPA \cite{Kolbe94} gives the inclusive rates of
0.342, 0.969, and 26.2 $\times 10^5$ s$^{-1}$ for $^{12}$C, $^{16}$O
and $^{40}$Ca; to be compared with the measured rates of
0.320, 0.924, and 25.6 $\times 10^5$ s$^{-1}$ for the same nuclei.
This good agreement is again obtained without any parameter adjustment.
In particular, as discussed in Ref.\cite{Kolbe94}, no renormalization of the
axial vector coupling constant $g_A$ in nuclear medium is required.

What are the momentum transfers and excitation energies
involved in the inclusive reactions which we would like
to describe? For the $^{12}$C($\nu_e,e$)$^{12}$N* with the
electron neutrinos originating in the muon decay at rest,
the typical momentum transfer is $\langle |\vec{q}| \rangle \simeq$
50 MeV, and the typical excitation energy
is $\omega \simeq$ 20 MeV. For the inclusive muon
capture $^{12}$C($\mu^-,\nu_{\mu}$)$^{12}$B* we have
$\langle |\vec{q}| \rangle\simeq$ 90 MeV
and the typical excitation energy
is $\omega \simeq$ 25 MeV. Finally for the
$^{12}$C($\nu_{\mu},\mu^-$)$^{12}$N*
with the muon neutrinos originating in the pion decay
in flight at LAMPF we have
$\langle |\vec{q}| \rangle\simeq$  200 MeV
and the typical excitation energy
is $\omega \simeq$ 40 MeV.
The excitation energies should be compared with the
nuclear shell spacing $\hbar \omega \simeq 41/A^{1/3}$ MeV, which for
$^{12}$C is equal to about 18 MeV. Thus, in order to describe
all the above inclusive processes in the framework of the nuclear
shell model, one would have to include fully and consistently at
least all $2 \hbar \omega$ excitations, and possibly even
the $3 \hbar \omega$ ones. This is not impossible, but
represents a formidable
task. On the other hand, the CRPA can easily handle such configuration
spaces. Moreover, it properly describes the continuum nature
of the final nucleus. Finally, as argued above, the crudeness with which
the correlations of the $p$ shell nucleons is treated in the CRPA
is expected to be relatively unimportant.

For the inclusive reaction $^{12}$C($\nu_e,e^-$)$^{12}$N$^{*}$,
with $\nu_e$ neutrinos from the muon decay-at-rest
the calculation gives\cite{Kolbe94}
the cross section of 6.3 $\times 10^{-42}$ cm$^2$
using the Bonn potential based G-matrix as the residual interaction,
and 5.9 $\times 10^{-42}$ cm$^2$ with the schematic Migdal force.
(The two different residual interactions are used so that
one can estimate the uncertainty associated with this aspect
of the problem.)
Both are clearly compatible with the measured values of
$6.4 \pm 1.45[stat] \pm 1.4[syst] \times 10^{-42}$ cm$^2$  by
the Karmen collaboration\cite{Karmen} (the more
recent result gives somewhat smaller value
$5.1 \pm 0.6 \pm 0.5$\cite{Maschuw}) and with
$5.7 \pm 0.6[stat] \pm 0.6[syst] \times 10^{-42}$ cm$^2$ obtained by
the LSND collaboration\cite{LSND} . If one wants to disregard
the error bars (naturally, one should not do that),
one can average the two calculated values
as well as the two most recent measurements and
perhaps conclude that the CRPA calculation
seems to exceed the measured values by about 10-15\%.
A similar tendency can be found, again with some
degree of imagination, in the comparison of the muon
capture rates discussed earlier.

So far we have found that the CRPA describes the
inclusive reactions quite well. Other theoretical calculations,
e.g.\cite{Oset98,Auerbach} describe these reactions
with equal success. This is no longer the case when we
consider the reaction $^{12}$C($\nu_\mu,\mu$)$^{12}$N$^{*}$
with the higher energy $\nu_{\mu}$ neutrinos from the pion
decay-in-flight. This reaction involves
larger momentum transfers and populates
states higher up in the continuum of  $^{12}$N.
The CRPA calculation\cite{Kolbe94,Kolbe95} gives the cross section of
19.2 $\times 10^{-40}$ cm$^2$,
considerably larger than the measured\cite{LSND} value of
$11.3 \pm 0.3[stat] \pm 1.8[syst]$ in the same units.
The origin of the discrepancy is not clear, but as stressed in the discussion
of the exclusive reaction, the $\nu_{\mu}$ flux normalization is not
a likely culprit. While Ref. \cite{Oset} confirms our result, Ref.\cite{Mintz}
gets a value close to the experiment by using a generalization of the
EPT approach. 

Other recent theoretical calculations span the region between
the CRPA and experiment. So, Singh et al.\cite{Oset98} give
$16.65 \pm 1.37$ $\times 10^{-40}$ cm$^2$, clearly higher
than the experiment but somewhat lower than the CRPA.
On the other hand, Ref.\cite{Auerbach} gives 13.5 - 15.2 in the
same units, a value which is even closer to the experiment.
The main difference in that work is the inclusion of pairing which 
is not expected to represent a substantial effect.

This discrepancy has been with us for quite some time now.
It clearly exceeds the 10-15\% perhaps suggested by the
lower energy inclusive reactions discussed above.
It would be very important to perform a large scale shell model
calculation, including up to 3$\hbar \omega$
excitations,  to put the matter to rest. Attempts to do that
are in Refs. \cite{Hayes01,Brown02}.

\section{Supernova neutrinos}

One of the most important application of the neutrino-nucleus interaction
is the detection of supernova neutrinos. In this section, after few
introductory remarks, we describe several examples of the calculated
charged and neutral current cross sections on oxygen, argon, iron, and lead.
All these nuclei are being considered (or actually are already
used) as targets for
the supernova neutrino detection. Some of these cross sections are amenable
to tests using the spallation neutron source, since the neutrino
spectra of stopped pions and muons are quite similar to the
expected neutrino spectra from the core collapse supernovae.
The general review of the field can be found in Ref.\cite{Raffelt}. 

Supernova neutrinos from SN1987a, presumably all $\bar{\nu}_e$,
had been observed by the Kamiokande and IMB detectors \cite{Hirata,IMB}
and have confirmed the general supernova picture.
However,
the supernova models predict distinct differences in the 
neutrino distributions for the various flavors and thus a more
restrictive test of the current supernova theory requires the abilities
of neutrino spectroscopy by the neutrino detectors.
Current (e.g. Superkamiokande, SNO, KamLAND)
and future detectors (including the proposed OMNIS \cite{Omnis}
and LAND \cite{Land} projects) have this  capability 
and will be able to distinguish between the different neutrino flavors
and determine their individual spectra. For the water \v{C}erenkov detectors
(SNO and Superkamiokande) $\nu_x$ neutrinos can be detected by specific
neutral-current events \cite{Langanke96,SNO}, 
while the OMNIS and LAND detectors are
proposed to detect neutrons spalled from target nuclei by charged- and
neutral-current neutrino interactions.

Theoretical models predict characteristic 
differences in the neutrino distributions for the various neutrino flavors
(so-called temperature hierarchy).
The $\mu$ and $\tau$ neutrinos and
their antiparticles (combined referred to as $\nu_x$)
decouple deepest in the star,
i.e. at the highest temperature, and have an average energy of 
$\bar{E}_{\nu} \simeq 25$ MeV. The $\nu_e$ and $\bar{\nu}_e$ neutrinos interact
with the neutron-rich matter via $\nu_e+n \rightarrow p + e^-$ and
$\bar{\nu}_e + p \rightarrow n + e^+$; the $\bar{\nu}_e$ neutrinos
have a higher average energy 
($\bar{E}_{\nu} \simeq  16$ MeV) than the $\nu_e$
neutrinos ($\bar{E}_{\nu} \simeq  11$ MeV). Clearly an observational
verification of this temperature hierarchy would establish a strong test
of the supernova models. 
The distribution of the various supernova neutrino species is usually
described by the pinched Fermi-Dirac spectrum
\begin{equation}
n(E_\nu) = \frac{1}{F_2({\alpha}) T^3} \frac{E_\nu^2}{exp[(E_\nu/T)-\alpha]+1}
\end{equation}
where $T,\alpha$ are parameters fitted to numerical spectra, and
$F_2(\alpha)$ normalizes the spectrum to unit flux. The transport
calculations of Janka \cite{Janka} yield spectra with $\alpha \sim 3$
for all neutrino species. While this choice also gives good fits to the
$\nu_e$ and ${\bar \nu}_e$ spectra calculated by Wilson and Mayle
\cite{Wilson}, their $\nu_x$ spectra favor $\alpha=0$. In the following
we will present results for charged- and neutral current reactions
on several target nuclei for both values of $\alpha$. In particular
we will include results for those 
($T,\alpha$) values which are currently favored 
for the various neutrino types (T in MeV):
$(T,\alpha)= (4,0)$ and (3,3) for $\nu_e$ neutrinos, (5,0) and (4,3) for
${\bar \nu}_e$ neutrinos and (8,0) and (6.26,3) for $\nu_x$ neutrinos.
However, it is worthwhile to point out that the degree of separation
in energy of the different flavors is somewhat model dependent, as
shown e.g. in Ref. \cite{Keil02}. It is therefore even more important
to determine the relevant parameters experimentally.

As stated above, it is usually  sufficient to
evaluate the various neutrino-induced reaction  cross sections  
within the 
RPA. However, the RPA often does not recover sufficient nucleon-nucleon
correlations to reliably reproduce the quenching and fragmentation of
the Gamow-Teller (GT) strength distribution in nuclei.
Therefore, the response of the $\lambda^\pi = 1^+$
operator should be evaluated on the basis 
of an interacting shell model, if such calculations are
feasible. While the double-magic nucleus $^{16}$O does not
allow GT excitations, shell model calculations in reliably large model
spaces are possible for $^{40}$Ar and $^{56}$Fe. For $^{208}$Pb 
GT transitions are Pauli-blocked for $({\bar \nu_e},e^+)$ reactions. The
modeling of GT transitions in the $(\nu_e,e^-)$ reactions on
$^{208}$Pb would require much too large model spaces;
these transitions must also be
evaluated within the RPA approach. 

In the following we will refer to
a `hybrid model' 
if the allowed transitions have been studied based on the
interacting shell model, while the forbidden transitions were calculated
within the random phase approximation. The studies for $^{40}$Ar and
$^{56}$Fe are performed in such a hybrid model, while the ones for
$^{16}$O and $^{208}$Pb are RPA calculations for all multipoles.

We note that the GT operator
corresponds to the $\lambda^\pi=1^+$ operator only in the
limit of momentum transfer $q \rightarrow 0$. As it has been pointed
out in \cite{Kolbe99,Hektor}, the consideration of the finite-momentum
transfer in the operator results in a reduction of the cross sections.
To account for the effect
of the finite momentum transfer we performed RPA calculations for
the $\lambda^\pi =1^+$ multipole operator at finite momentum transfer
$q$ (i.e. $\lambda(q)$) and for $q=0$ (i.e., $\lambda(q=0)$) and 
scaled the shell model GT strength distribution by the ratio of
$\lambda(q)$ and $\lambda(q=0)$ RPA cross sections. 
The correction is rather small for $\nu_e$ neutrinos stemming
from muon-decay-at-rest neutrinos (e.g., for LSND and Karmen) 
or for supernova $\nu_e$ neutrinos. The correction is, however, sizeable
for higher neutrino energies.

\subsection{Oxygen}

Observation of neutrinos from the
SN1987A did not allow to test in detail the neutrino
distribution and, in particular, it gave no information about $\nu_x$
neutrinos which, as we discussed above, should decouple deepest in the
star. 
The observability of supernova 
neutrinos has significantly improved since
the Superkamiokande (SK) detector, with a threshold of 5 MeV
and with 30 times the size of Kamiokande,
became operational \cite{Totsuka}.

\begin{figure}[htb]
\begin{center}
  \leavevmode
   \includegraphics[width=0.5\columnwidth,height=8.5cm]{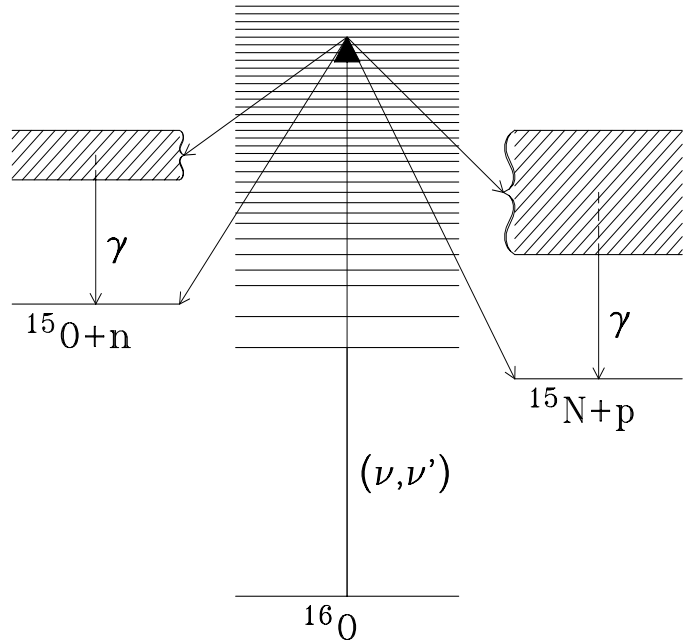}
\caption{Schematic illustration of the dection scheme for the neutral
current detection in water \v{C}erenkov detectors.}
\label{fig:ncscheme}
\end{center}
\end{figure}

\begin{figure}[htb]
\begin{center}
  \leavevmode
   \includegraphics[width=0.8\columnwidth,height=12.5cm]{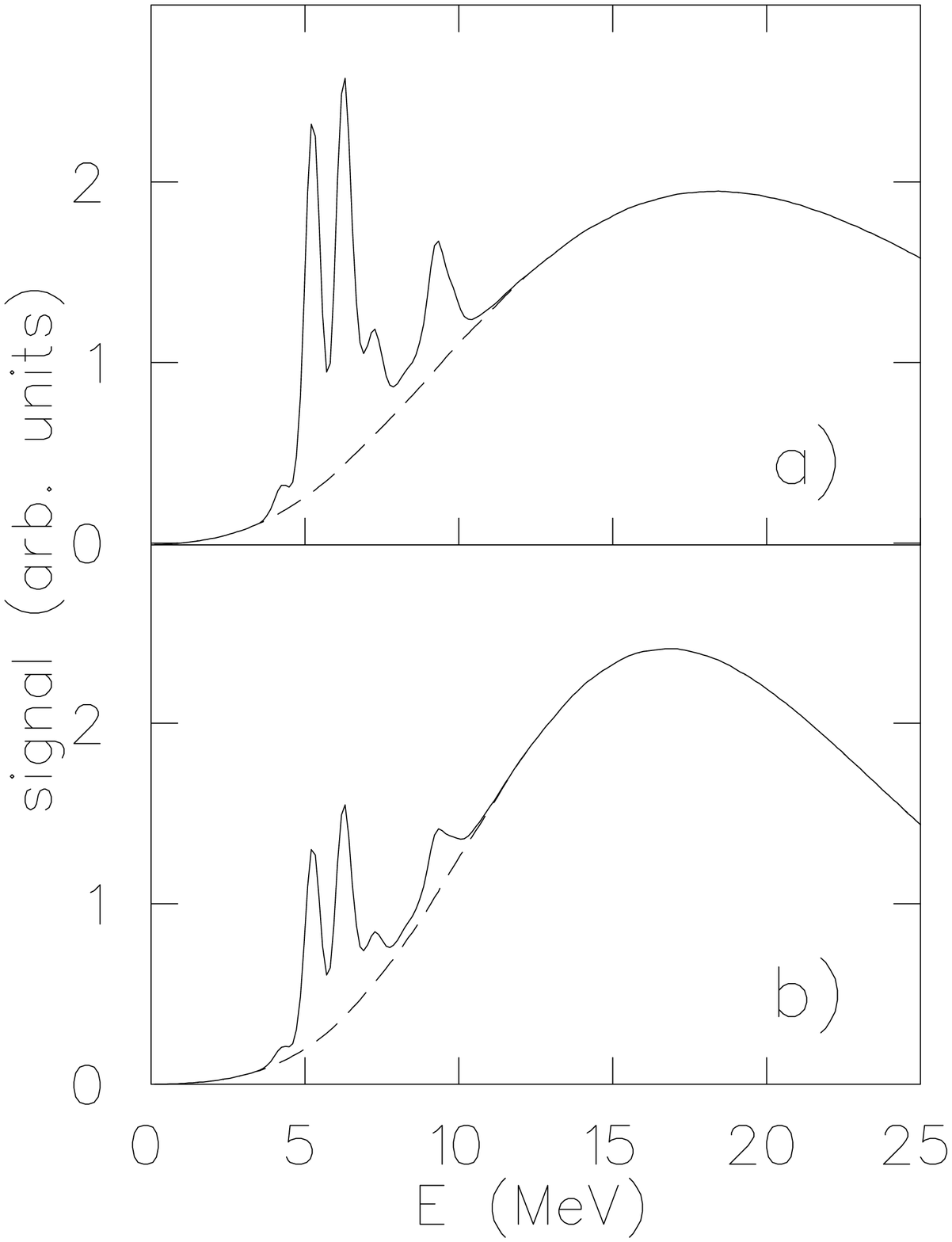}
\caption{
Signal expected from supernova neutrinos in a water \v{C}erenkov detector
calculated for two different types of neutrino distributions (without
chemical potential (above) and with chemical potential $\alpha=3T$ and
temperatures $T=6.26 $ MeV (for $\nu_x$) and $T=4$ MeV (for ${\bar
\nu}_e$)). The bulk of the signal stems from ${\bar \nu}_e$ neutrinos
reacting with protons, while the $\nu_x$ neutrinos induce the superimposed
signal at energies $E=5-10$ MeV (from \protect\cite{Langanke96}).
}
\label{fig:kamio}
\end{center}
\end{figure}

Clearly, many of the primary neutral- and charged-current 
$\nu$-induced reactions in SK occur on $^{16}$O. 
Atmospheric neutrinos and also supernova $\nu_x$ neutrinos
have high enough energies so that the final
nucleus in the primary reaction will be in an excited state which will
then decay by a cascade of particle and $\gamma$ emissions.
This fact has  been used to propose
a signal for the observation of $\nu_x$ neutrinos 
in water \v{C}erenkov detectors 
\cite{Langanke96}.
Schematically the detection scheme works
as follows (see Fig. \ref{fig:ncscheme}). 
Supernova $\nu_x$ neutrinos, with average energies
of $\approx 25$ MeV, will predominantly excite $1^-$ and $2^-$ giant
resonances in $^{16}$O via  the 
$^{16}$O($\nu_x,\nu^\prime_x$)$^{16}$O$^\star$ neutral current reaction
\cite{Kolbe92}. These resonances are above the particle thresholds and
will mainly decay by proton and neutron emission. 
Although these
decays will be dominantly to the ground states of
$^{15}$N and $^{15}$O, respectively, some of them will go to
excited states in these nuclei. In turn, if these excited states are below
the particle thresholds in 
$^{15}$N ($E^\star < 10.2$ MeV) or
$^{15}$O ($E^\star < 7.3$ MeV), 
they will decay by $\gamma$ emission.
As the first excited states in both of these mirror nuclei 
($E^\star =5.27$ MeV in $^{15}$N and
$E^\star = 5.18$ MeV in $^{15}$O) 
are at energies larger than the SK detection threshold, all of the 
bound excited
states in $^{15}$N and $^{15}$O below 
will emit photons which can be observed in SK.

Based on a calculation which 
combines the
Continuum RPA with the statistical model 
\cite{Langanke96}, 
Superkamiokande is expected to observe about 700
$\gamma$ events in the energy window $E=5-10$ MeV, induced by
$\nu_x$ neutrinos (with a FD distribution of $T=8$ MeV),
for a supernova going off at 10 kpc ($\approx 3 \cdot 10^4$ light years
or the distance to the galactic center), see Fig.\ref{fig:kamio}.
This is to be compared
with a smooth background of about 270 positron events from the
 ${\bar \nu_e} + p \rightarrow n + e^+$ reaction in the same 
energy window.  
The number of
events produced by supernova $\nu_x$ neutrinos via the scheme proposed here
is larger than the total number of events 
expected from $\nu_x$-electron scattering (about 80
events \cite{Totsuka}). More importantly, the $\gamma$ signal
can be unambiguously identified from the observed spectrum in the SK detector,
in contrast to the more difficult identification from $\nu_x$-electron
scattering. 

The cascades of decays, following the inelastic excitation of $^{16}$O
by atmospheric or supernova $\nu_x$ neutrinos,
can also result in the production of
$\beta$-unstable nuclei. If the $Q_\beta$ values of these nuclei are
above the observational threshold energy in SK ($\sim 5$ MeV),
these decays might be detectable, and since they are usually delayed,
might offer an additional characteristic signature of the
neutrino-induced reactions. As possible candidates  
$^{16}$N, $^{15}$C,
$^{12}$B and $^{12}$N have been identified \cite{Nussinov,Kolbe02}.

\begin{figure}[ht]
  \begin{center}
    \leavevmode
    \includegraphics[width=0.7\columnwidth,height=10.5cm]{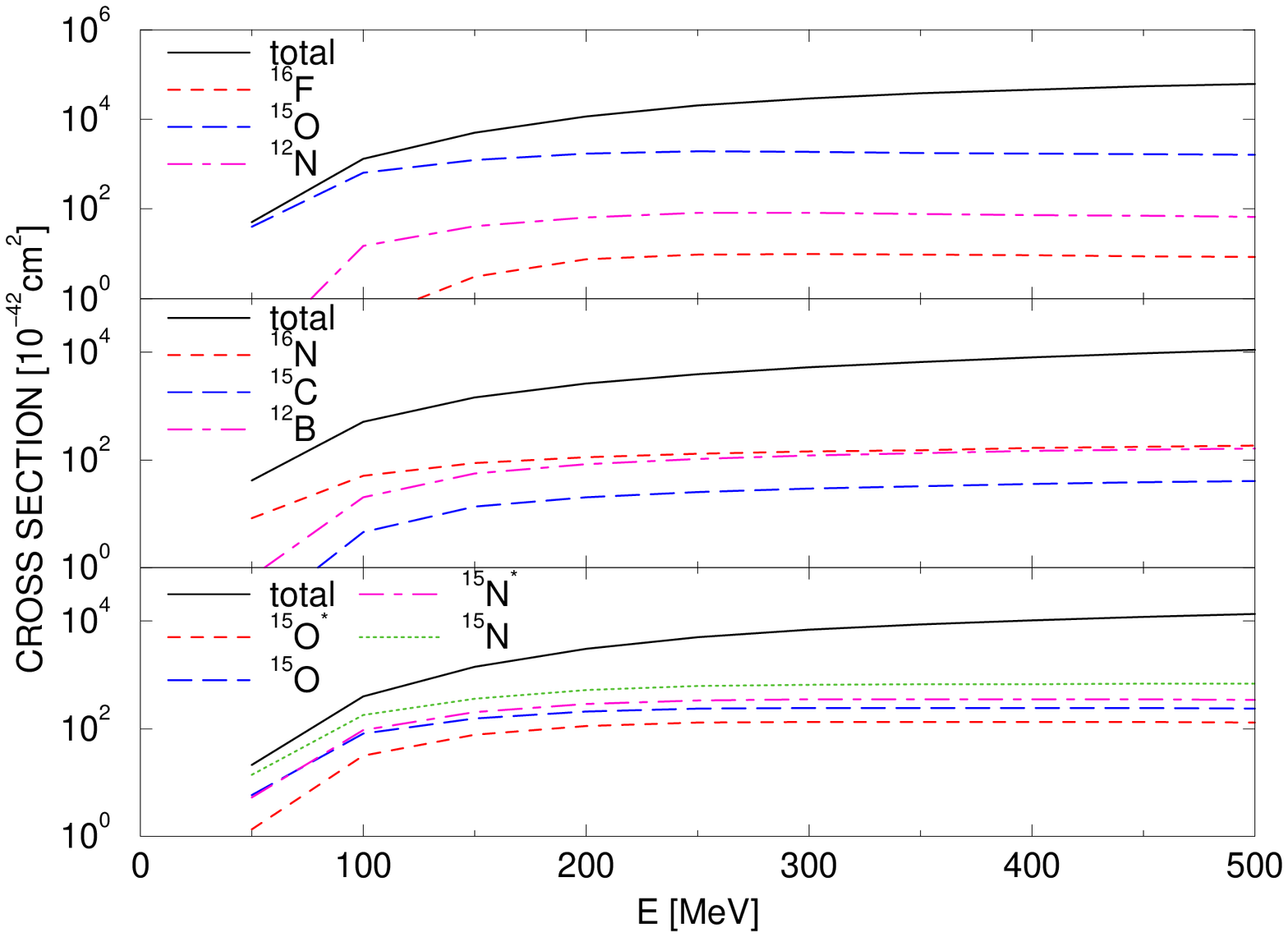}
    \caption{Total and partial cross sections to selected $\beta$-unstable 
nuclear ground states
for $(\nu_e,e^-)$ (upper),
$({\bar \nu_e},e^+)$ (middle) and $(\nu,\nu')$ (lower part)  
reactions on $^{16}$O as function of neutrino energy.
As explained in the text, the symbols $^{15}$O$^*$ and $^{15}$N$^*$
in the lower panel denote the partial cross sections leading through
particle-bound excited states in these nuclei. 
        }
    \label{fig_nuss}    
  \end{center}
\end{figure}

Fig. \ref{fig_nuss} shows the partial $({\bar \nu_e},e^+)$ 
and $(\nu_e,e^-$) cross sections
leading to
the $\beta$-unstable $^{16}$N, $^{15}$C and $^{12}$B 
and $^{16}$F, $^{15}$O and $^{12}$N
ground states in the final channel;
these cross sections reflect the sum over all cross sections with
particle-bound states of these nuclei as these excited states will fast
decay to the ground state by $\gamma$ emission. 
Our calculations have
been performed up to neutrino energies $E_\nu=500$ MeV. At higher
energies the total cross sections can be obtained from a relativistic
Fermi gas model \cite{Engel}, 
including, however,  additional channels like pion production.

The results make it possible 
to draw some interesting conclusions. While
the total cross sections increase with neutrino energies, most of
this increase
goes into new channels which
open up with increasing neutrino energy and
the partial
cross sections leading to definite states have the tendency to saturate.
Thus our partial cross sections to these states, obtained for, say,
$E_\nu=500$ MeV, can be used to derive upper limits for the 
corresponding branching ratios 
expected for atmospheric neutrinos which, on average,
have even larger energies. The total and selected partial cross sections
for the neutral- and charged-current reactions on $^{16}$O for different
supernova neutrino spectra are summarized in 
Tables \ref{tab_nuss1}-\ref{tab_nuss3}.  

A detailed discussion of the various cross sections is given in
\cite{Kolbe02}.
In general, these RPA cross sections are significantly smaller than
estimated in \cite{Nussinov}. In particular, they indicate that,
for typical atmospheric neutrino
energies,  the
partial cross sections leading to $\beta$-unstable nuclei
is, unfortunately, too small so that the observation of the decay of
these nuclei does not constitute an additional viable signal for
neutrino-induced reactions 
in Superkamiokande. 

For supernova $\bar{\nu}_e$ neutrinos, however, the
$^{16}$O$({\bar \nu_e}, e^+$)$^{16}$N reaction can
produce an observable additional signal in Superkamiokande for
supernovae from within our galaxy.
The reaction leads to
excited states in $^{16}$N at rather low  energies, which then
dominantly decay by neutron emission. However, 
as can be seen in Table \ref{tab_nuss3},
a sizable fraction
of the $({\bar \nu_e},e^+)$
reactions also excite  particle-bound states in
$^{16}$N, followed then by the $\beta$ decay of the $^{16}$N ground
state. Assuming the standard antineutrino supernova spectrum (with $T=5$ MeV)
the partial cross section of $3.5 \cdot 10^{-43}$ cm$^2$
corresponds to about 40
supernova ${\bar \nu_e}$-induced events in SK leading to the $^{16}$N ground
state and which can be identified by the delayed $\beta$ decay
for a hypothetical supernova in the galactic center. Note,
however, that this event rate corresponds to less than $1\%$ of the
total supernova neutrino rate in SK, with positrons being produced
by the ${\bar \nu_e} + p
\rightarrow e^+ + n$ reaction giving the dominating signal. Thus, it
is unlikely that $\beta$ decays from $^{16}$N, generated by
$^{16}$O$({\bar \nu_e}, e^+$)$^{16}$N in the supernova 1987A, were observed
by the Kamiokande detector. 
The partial 
$^{16}$O$({\bar \nu_e}, e^+$)$^{16}$N 
reaction cross section increases by more than a factor of 6,
if complete ${\bar \nu_e} \leftrightarrow
{\bar \nu_{\mu,\tau}}$ oscillations occur, constituting then a very
sizable and clean signal for SK.

\begin{table}[thb]
\caption{Partial cross sections for neutral-current  
           neutrino-induced reactions on $^{16}$O. 
A Fermi-Dirac distribution with $T=8$ MeV and zero chemical potential,
which is  typical for supernova
$\nu_\mu$ and $\nu_\tau$ neutrinos and their antiparticles,
has been assumed. The cross section, in units of $10^{-42} {\rm cm}^2$,
 represents the
average for neutrino and antineutrino reactions, 
and the  exponents are given in parentheses. The asterisks indicate
that the cross sections have been summed over all particle-bound states.}
  \begin{center}
    \begin{tabular}{|l|c|}   
      \rule[-1ex]{0em}{4ex} neutrino reaction &  partial $\sigma$ \\ 
       \hline 
       total      &  5.19 (00)  \\
       \nucleus{\rm O}{16}($\nu,\nu' \gamma$)\nucleus{\rm O^*}{16} &
         3.19 ( -3) \\ 
       \nucleus{\rm O}{16}($\nu,\nu'$ n)\nucleus{\rm O}{15}(gs) &
         9.73 (-1) \\ 
       \nucleus{\rm O}{16}($\nu,\nu'$ p)\nucleus{\rm N}{15}(gs) &
         1.85 ( 00) \\ 
       \nucleus{\rm O}{16}($\nu,\nu' n\gamma$)\nucleus{\rm O^*}{15} &
         3.48 ( -1) \\ 
       \nucleus{\rm O}{16}($\nu,\nu'$ nn)\nucleus{\rm O^*}{14} &
         6.11 ( -3) \\ 
       \nucleus{\rm O}{16}($\nu,\nu'$ np)\nucleus{\rm N^*}{14} &
         4.40 ( -1) \\ 
       \nucleus{\rm O}{16}($\nu,\nu' p\gamma$)\nucleus{\rm N^*}{15} &
         1.29 ( 00) \\ 
       \nucleus{\rm O}{16}($\nu,\nu'$ pp)\nucleus{\rm C^*}{14} &
         8.35 ( -2) \\ 
       \nucleus{\rm O}{16}($\nu,\nu' p\alpha$)\nucleus{\rm B^*}{11} &
         9.15 ( -2) \\ 
       \nucleus{\rm O}{16}($\nu,\nu' n\alpha$)\nucleus{\rm C^*}{11} &
         3.88 ( -2) \\ 
    \end{tabular}
  \end{center}
  \label{tab_nuss1}
\end{table}

\begin{table}[thb]
\caption{Partial cross sections for charged-current  
           neutrino-induced reactions on $^{16}$O. 
Fermi-Dirac distributions with $T=4$ MeV and $T=8$ MeV 
and zero chemical potential have been assumed. The first
is  typical for supernova
$\nu_e$ neutrinos, while the second can occur for complete $\nu_e
\leftrightarrow \nu_\mu$ oscillations.
           The 
           cross sections are given in units of $10^{-42} {\rm cm}^2$,
           exponents are given in parentheses.}
  \begin{center}
    \begin{tabular}{|l|c|c|}   
      \rule[-1ex]{0em}{4ex} neutrino reaction  &  $\sigma$, $T=4$
MeV &
 $\sigma$, $T=8$ MeV \\ 
\hline
       total & 1.91 (-1) & 1.37 (+1) \\                  
       \nucleus{\rm O}{16}($\nu, e^-$ p $\gamma$)\nucleus{\rm O}{15}(gs) &
         1.21 (-1)  &  6.37 (00) \\ 
       \nucleus{\rm O}{16}($\nu, e^-$ p $\gamma$)\nucleus{\rm O^*}{15} &
         4.07 (-2)  &  3.19 (00) \\ 
       \nucleus{\rm O}{16}($\nu, e^-$ np)\nucleus{\rm O^*}{14} &
         3.92 (-4)  &  1.76 (-1) \\ 
       \nucleus{\rm O}{16}($\nu, e^-$ pp)\nucleus{\rm N^*}{14} &
         2.61 (-2)  &  3.26 (00) \\
       \nucleus{\rm O}{16}($\nu, e^- \alpha$)\nucleus{\rm N^*}{12} &
         1.16 (-3)  &  1.31 (-1) \\ 
       \nucleus{\rm O}{16}($\nu, e^- p\alpha$)\nucleus{\rm C^*}{11} &
         1.55 (-3)  &  5.66 (-1) \\ 
       \nucleus{\rm O}{16}($\nu, e^- pn\alpha$)\nucleus{\rm C^*}{10} &
         1.11 (-6)  &  3.28 (-3) \\ 
    \end{tabular}
  \end{center}
  \label{tab_nuss2}
\end{table}

\begin{table}[thb]
\caption{Partial cross sections for charged-current  
           antineutrino-induced reactions on $^{16}$O. 
Fermi-Dirac distributions with $T=5$ MeV and $T=8$ MeV 
and zero chemical potential have been assumed. The first
is  typical for supernova
${\bar \nu_e}$ neutrinos, while the second can occur for complete 
${\bar \nu_e}
\leftrightarrow {\bar \nu_\mu}$ oscillations.
           The 
           cross sections are given in units of $10^{-42} {\rm cm}^2$,
           exponents are given in parentheses.}
  \begin{center}
    \begin{tabular}{|l|c|c|}   
      \rule[-1ex]{0em}{4ex} neutrino reaction  &  $\sigma$, $T=5$
MeV &
 $\sigma$, $T=8$ MeV \\ 
\hline
       total & 1.05 (00) & 9.63 (00) \\                  
       \nucleus{\rm O}{16}(${\bar \nu}, e^+$ )\nucleus{\rm N}{16}(gs) 
       &  3.47 (-1)  &  2.15 (00) \\ 
       \nucleus{\rm O}{16}(${\bar \nu}, e^+$ n )\nucleus{\rm N}{15}(gs) 
       &  5.24 (-1)  &  4.81 (00) \\ 
       \nucleus{\rm O}{16}(${\bar \nu}, e^+$ n $\gamma$)\nucleus{\rm N^*}{15} 
       &  1.47 (-1)  &  1.90 (00) \\ 
       \nucleus{\rm O}{16}(${\bar \nu}, e^+$ np)\nucleus{\rm C^*}{14} 
       &  4.56 (-3)  &  1.38 (-1) \\ 
       \nucleus{\rm O}{16}(${\bar \nu}, e^+$ nn)\nucleus{\rm N^*}{14} 
       &  5.50 (-3)  &  1.81 (-1) \\
       \nucleus{\rm O}{16}(${\bar \nu}, e^+ \alpha$)\nucleus{\rm B^*}{12} 
       &  1.07 (-2)  &  1.91 (-1) \\ 
       \nucleus{\rm O}{16}(${\bar \nu}, e^+ n\alpha$)\nucleus{\rm B^*}{11} 
       &  6.20 (-3)  &  2.16 (-1) \\ 
    \end{tabular}
  \end{center}
  \label{tab_nuss3}
\end{table}

Another interesting reaction, leading to an observable $\beta$ decay in
SK, is $^{16}$O(${\bar \nu_e},e^+ \alpha)$$^{12}$B. 
For the standard spectrum we find a partial cross section of $1.1 \cdot
10^{-44}$ cm$^{2}$, which increases to $1.9 \cdot 10^{-43}$ cm$^2$ for
the case of oscillations. This translates into $O(10)$ $^{12}$B
decays in the SK detector for a 
supernova in the galactic center at 10 kpc and if complete 
${\bar \nu_e} \leftrightarrow
{\bar \nu_{\mu,\tau}}$ oscillations occur.

\subsection{Argon}

The proposed ICARUS detector uses a liquid Argon technique
providing energy and direction measurements for
electrons, muons, pions and protons \cite{Icarus}. 
Observation of  supernova neutrinos is also
anticipated in the detector.

\begin{figure}[ht]
  \begin{center}
    \leavevmode
    \includegraphics[width=0.7\columnwidth,height=6.5cm]{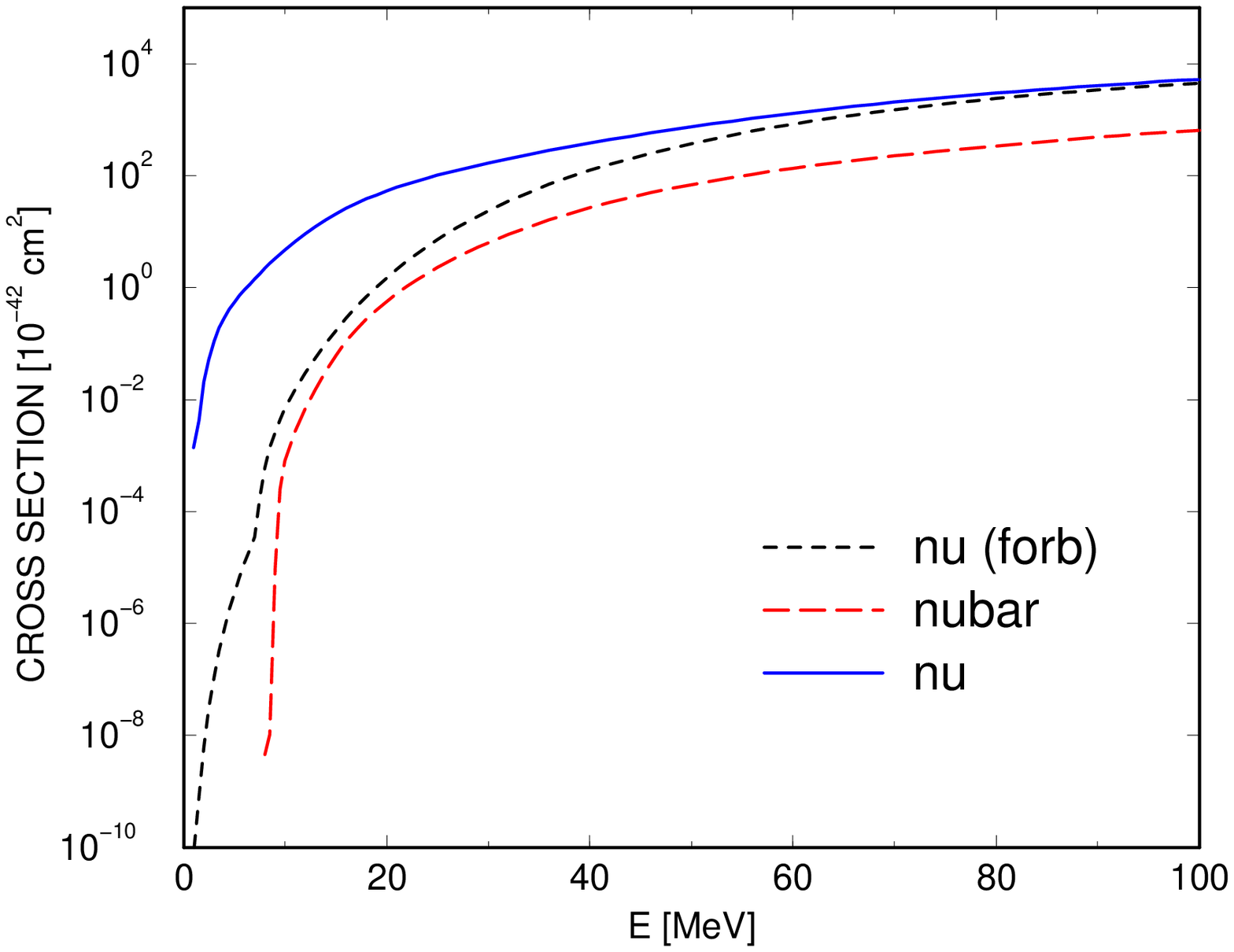}
    \caption{Total $(\nu_e,e^-)$ (solid) and $({\bar \nu_e},e^+)$
(dashed) cross sections on $^{40}$Ar, calculated within the RPA approach.
The short-dashed line shows the forbidden contributions to the
    $(\nu_e,e^-)$ cross sections.
        }
    \label{fig_argon}    
  \end{center}
\end{figure}

The detection and analysis of supernova neutrinos by the ICARUS detector
requires the knowledge of the neutrino-induced cross sections on
$^{40}$Ar for neutrinos and antineutrinos with energies up to about 100 MeV.
At low neutrino energies the $(\nu_e,e^-)$ reaction is dominated by
allowed GT transitions. However, the calculation of this cross section
constitutes a quite challenging nuclear structure problem. In the
Independent Particle Model (IPM),  $^{40}$Ar
($Z=18$, $N=22$) corresponds to a 2-hole proton configuration in the
$sd$ shell and a 2-particle neutron configuration in the $pf$ shell.
One thus has to expect that cross-shell correlations will strongly
influence the structure of the low-lying states in $^{40}$Ar, including
the GT$_-$ response of the ground state. To describe such correlations
requires shell model studies in the complete $sd-pf$ shell which are
currently not feasible by diagonalization methods. One therefore has
to  rely on truncated shell model calculations in which only a
restricted number of particles are allowed to be excited from the $sd$
shell to the $pf$ shell.
The best study to date has been performed by Ormand \cite{Ormand} who
calculated the low-lying GT response of the $^{40}$Ar ground state
within a $2 \hbar \omega$ shell model diagonalization in which he
considered 2-particle excitations from the $sd$ into the $pf$ shell,
carefully avoiding spurious center-of-mass excitations. His calculation
is appropriate for solar neutrinos with energies $E_\nu \le 14$ MeV.
Importantly, the shell model study showed that the GT transitions
dominate over the allowed Fermi transition to the IAS in $^{40}$K at
4.38 MeV. However, due to the (rather severe) truncation of the model
space this shell model calculation violates the Ikeda sum rule and hence
misses GT strength at higher excitation energies in $^{40}$K. This
strength will be important for neutrinos (like those from a supernova)
which have larger energies than solar neutrinos and can excite the
daughter nucleus at higher energies. An appropriate shell model
calculation which describes this GT strength at higher energies is
currently not available. However, RPA studies of the
$^{40}$Ar($\nu_e,e^-)$$^{40}$K reaction have been performed considering
allowed and forbidden multipoles up to $J=4$. The respective cross
sections are shown in Fig. \ref{fig_argon}. It is evident that GT
transitions dominate the $(\nu_e,e^-)$ cross sections for neutrino
energies $E_\nu < 50$ MeV; at higher energies forbidden
(in particular spin-dipole) transitions cannot be neglected.

In the IPM, the GT$_+$ strength for $^{40}$Ar vanishes as all
GT transitions, in which a proton is changed into a neutron, are
Pauli-blocked. Although cross-shell
correlations might introduce a non-vanishing GT$_+$ strength, it should
be small. Hence it is reasonable to calculate the $({\bar \nu_e},e^+$)
cross section on $^{40}$Ar within the RPA approach. The obtained
results are also shown in Fig. \ref{fig_argon}.

\subsection{Iron}

The KARMEN collaboration used its  
sensitivity to the $^{56}$Fe($\nu_e,e^-$)$^{56}$Co background events 
to determine 
the cross section for this reaction  
for the DAR neutrino spectrum and obtained
$\sigma=(2.56 \pm 1.08 \pm 
0.43) \cdot 10^{-40}$ cm$^2$ \cite{Maschuw}. 
We calculate a result in close agreement,
$\sigma = 2.4 \cdot 10^{-40}$ cm$^2$ \cite{Kolbe99}.

We  extended
this investigation to the study of the charged- and neutral current reactions
on $^{56}$Fe. To allow also for the exploration of
potential oscillation scenarios we also  evaluated the cross sections
and the knockout neutron yields for various supernova neutrino spectra.
Table \ref{tab:fe1} 
summarizes the total and partial cross sections for neutral
current reactions on $^{56}$Fe. 
For $^{56}$Fe the neutron and proton thresholds open at 11.2 MeV and 10.18
MeV, respectively. But despite the slightly higher threshold energy, the
additional Coulomb repulsion in the proton threshold makes the
neutron channel  the dominating decay mode.  
The total and partial cross sections for charged current 
$(\nu_e,e^-)$ and (${\bar \nu}_e,e^+)$ reactions on
$^{56}$Fe  are listed in Table \ref{tab:fe2}. 
For the standard supernova $\nu_e$ spectrum 
($(T,\alpha)=(4,0)$),
the low-energy excitation spectrum is relatively strongly weighted
by phase space. Hence, in that case the 
$\nu_e$-induced reaction on $^{56}$Fe leads
dominantly to particle-bound states ($\sim 60\%)$ and therefore decays
by $\gamma$ emission. 

\begin{figure}[ht]
  \begin{center}
    \leavevmode
    \includegraphics[width=0.6\columnwidth,height=10.0cm,angle=270]{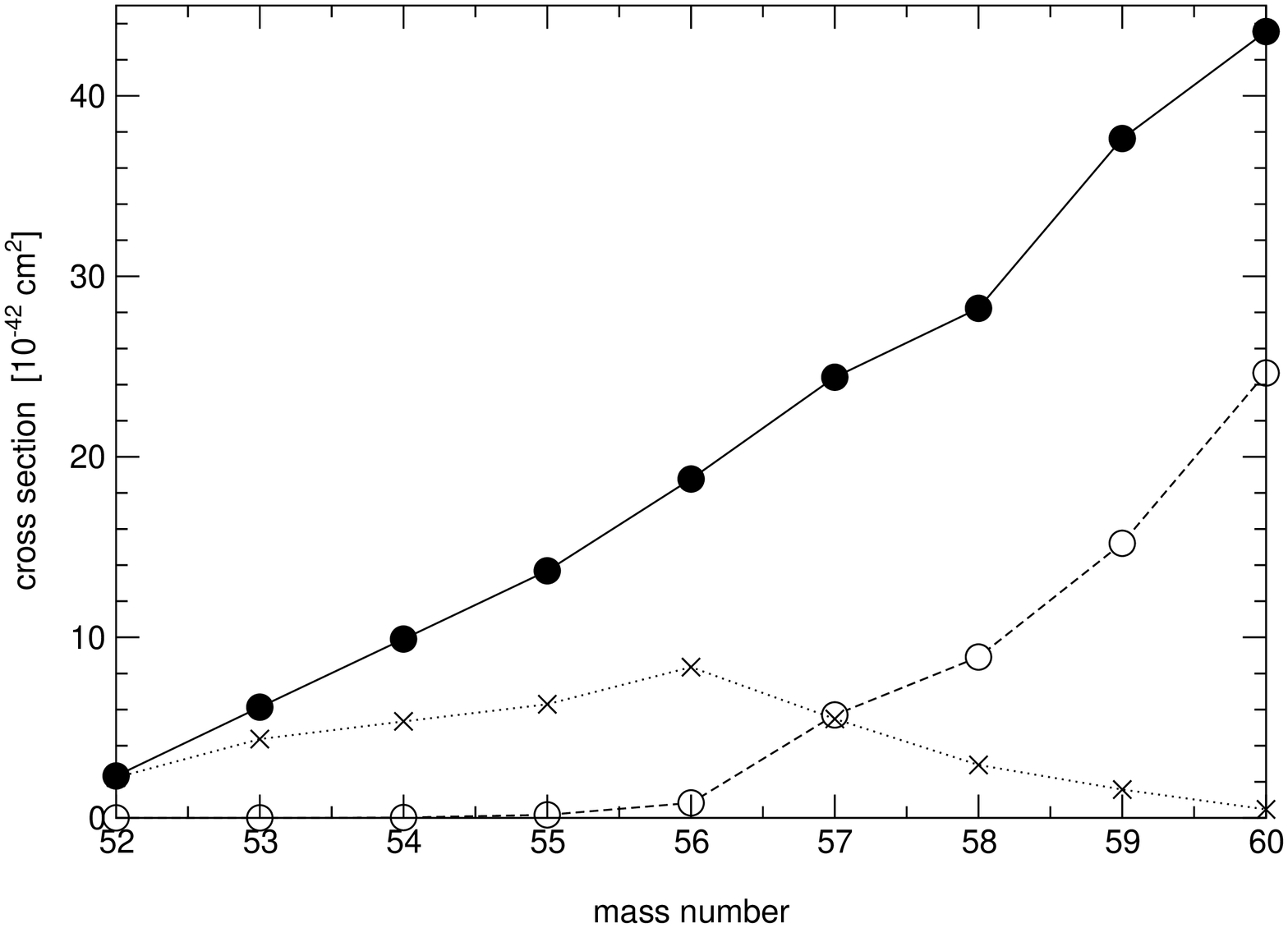}
\caption
{Total charged current cross section
for $\nu_e$ on $^{52-60}$Fe (filled circles connected by
the full line) and the partial neutron (empty circles, dashed line)
and proton (crosses, dotted line) spallation cross sections.
The difference between the total and the two partial cross
sections gives the gamma emission cross section
since the $\alpha$ channel is negligible.
The $\nu_e$ spectrum with $T$ = 4 MeV and $\alpha$ = 0
was used.}
\label{fig:toiv1}
\end{center}
\end{figure}

\begin{figure}[ht]
  \begin{center}
    \leavevmode
    \includegraphics[width=0.6\columnwidth,height=10.0cm,angle=270]{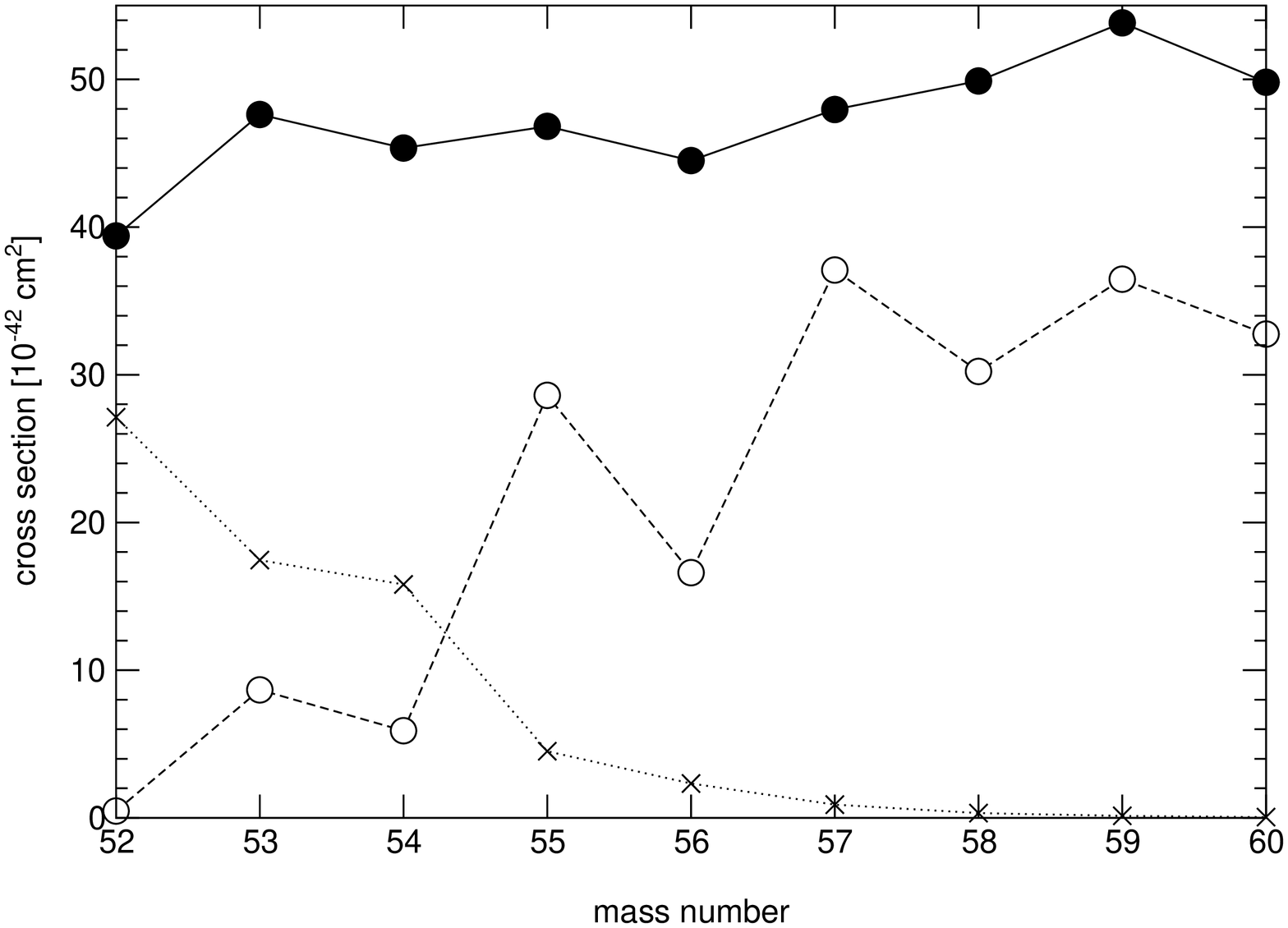}
\caption
{Total neutral current cross section
for $\nu_x$ on $^{52-60}$Fe averaged
between $\nu$ and $\bar{\nu}$ (filled circles connected by
the full line) and the partial neutron (empty circles, dashed line)
and proton (crosses, dotted line) spallation cross sections.
The difference between the total and the two partial cross
sections gives the gamma emission cross section
since the $\alpha$ channel is negligible.
The $\nu_x$ spectra with $T$ = 8 MeV and $\alpha$ = 0
were used.}
\label{fig:toiv2}
\end{center}
\end{figure}

The work of Ref. \cite{Kolbe99} has been extended to other iron
isotopes,
$^{52-60}$Fe, by Toivanen {\it et al.} \cite{Toivanen}. The respective
total charged-current $(\nu,e^-)$ and neutral-current $(\nu,\nu')$ cross
sections for typical supernova neutrino spectra (i.e. $T=4$ MeV for
$\nu_e$ neutrinos and $T=8$ MeV for $\nu_x$ neutrinos) are shown in Figs.
\ref{fig:toiv1} and \ref{fig:toiv2}. Additionally these figures show the
partial cross sections for the decay into the proton and neutron
channels. 
Since for the isotopes $^{52-56}$Co the proton threshold is lower
than the neutron threshold, the preferred decay mode for the
$(\nu_e,e^-)$ reaction on $^{52-56}$Fe is by proton
emission.
This is reversed for $^{58-60}$Co, where decay into    
the neutron channel is
preferred. Because parts of the Gamow-Teller and the
Fermi strengths are located
below the particle thresholds, these states decay by gamma emission
which
accounts basically for the difference between the total cross section
and the sum of the partial proton and neutron decay cross sections.
Decays into the
$\alpha$ channel are unfavored.

For the neutral-current reactions decay by proton emission is
favored in the proton-rich nuclei $^{52-54}$Fe. However, as the neutron
threshold decreases with increasing mass number along the isotope chain,
while the proton threshold energy increases,
the probability for decay into the neutron channel increases at the
expense of decay by proton emission. This is clearly reflected in the
trend of the partial cross sections as function of neutron excess.
Due to pairing the neutron threshold is higher in even-even
nuclei than in odd-A, explaining the odd-even staggering in the partial
neutron decay cross sections. Again, the difference of the total cross
sections compared to the sum of partial proton and neutron decay cross
sections gives the $(\nu,\nu'\gamma)$ cross sections,
caused mainly by the Gamow-Teller strength below the particle
thresholds.

We note that neutrino-induced reactions on nuclei in the iron mass
region might also play a role during the collapse and the shock-revival
phase of a supernova \cite{Haxton88,rmp}. Under these conditions, the
cross sections have to be evaluated at the finite temperature of the stellar
environment involving excited states of the parent nucleus. Relevant
cross sections are presented in \cite{Langanke01a,Sampaio02}.

\begin{table}[htb]
  \caption{Total cross sections for neutral current
                    neutrino scattering on $^{56}$Fe 
for different neutrino energy
                    spectra represented as Fermi-Dirac distributions. The
                    cross sections are in units of
                    $10^{-42} {\rm cm}^2$ and are averaged over neutrinos and
                    antineutrinos.}

   \begin{center}
   \begin{tabular}{|l|c|c|c|c|c|c|c|}
      \rule[-1ex]{0em}{4ex} { \hspace{1.8em}} ($T, \alpha$) & 
       (4,0) & (6,0) & (8,0) & (10,0) & (3,-3) & (4,-3) & (6.26,-3) \\ 
       \hline \hline 
       \nucleus{\rm Fe}{56}($\nu,\nu^{\prime} \gamma$)\nucleus{\rm Fe}{56} &
        2.5 ( 0)& 9.8 ( 0)& 1.7 ( 1)& 2.8 ( 1)& 1.2 ( 0)& 4.4 ( 0)& 1.6 ( 1)\\
       \nucleus{\rm Fe}{56}($\nu,\nu^{\prime}$n)\nucleus{\rm Fe}{55} &
        8.9 (-1)& 6.7 ( 0)& 2.2 ( 1)& 5.0 ( 1)& 2.8 (-1)& 1.7 ( 0)& 1.4 ( 1)\\
       \nucleus{\rm Fe}{56}($\nu,\nu^{\prime}$p)\nucleus{\rm Mn}{55} &
        1.2 (-1)& 1.0 ( 0)& 3.6 ( 0)& 9.3 ( 0)& 3.4 (-2)& 2.3 (-1)& 2.2 ( 0)\\
       \nucleus{\rm Fe}{56}($\nu,\nu^{\prime} \alpha$)\nucleus{\rm Cr}{52} &
        2.4 (-2)& 1.9 (-1)& 6.6 (-1)& 1.7 ( 0)& 6.4 (-3)& 4.4 (-2)& 4.0 (-1)\\
       \nucleus{\rm Fe}{56}($\nu,\nu^{\prime}$)X &
        3.6 ( 0)& 1.8 ( 1)& 4.3 ( 1)& 8.9 ( 1)& 1.5 ( 0)& 6.3 ( 0)& 3.3 ( 1)\\
    \end{tabular}
  \end{center}
 \label{tab:fe1}       
\end{table}

\begin{table}[htb]
  \caption{Total cross sections for charged current
                  neutrino scattering on $^{56}$Fe 
for different neutrino energy
                  spectra represented as Fermi-Dirac distributions. The
                  cross sections are in units of $10^{-42} {\rm cm}^2$.}
  \begin{center}
    \begin{tabular}{|l|c|c|c|c|c|c|c|} 
      \rule[-1ex]{0em}{4ex} { \hspace{1.8em}} ($T, \alpha$) &  
       (4,0) & (6,0) & (8,0) & (10,0) & (3,-3) & (4,-3) & (6.26,-3) \\
       \hline \hline
       \nucleus{\rm Fe}{56}($\nu_e,e^-\gamma$)\nucleus{\rm Co}{56} &
         9.8 ( 0)& 3.1 ( 1)& 6.1 ( 1)& 1.3 ( 2)& 7.7 ( 0)& 2.1 ( 1)& 7.5 ( 1)\\
       \nucleus{\rm Fe}{56}($\nu_e,e^-$n)\nucleus{\rm Co}{55} &
         7.5 (-1)& 8.0 ( 0)& 3.2 ( 1)& 8.1 ( 1)& 2.5 (-1)& 1.7 ( 0)& 2.0 ( 1)\\
       \nucleus{\rm Fe}{56}($\nu_e,e^-$p)\nucleus{\rm Fe}{55} &
         5.4 ( 0)& 3.2 ( 1)& 9.7 ( 1)& 1.7 ( 2)& 9.2 (-1)& 5.1 ( 0)& 4.7 ( 1)\\
       \nucleus{\rm Fe}{56}($\nu_e,e^-\alpha$)\nucleus{\rm Mn}{52} &
         6.1 (-2)& 9.7 (-1)& 4.8 ( 0)& 1.5 ( 1)& 3.0 (-2)& 2.1 (-1)& 2.9 ( 0)\\
       \nucleus{\rm Fe}{56}($\nu_e,e^-$)X &
         1.6 ( 1)& 7.2 ( 1)& 1.9 ( 2)& 4.0 ( 2)& 8.9 ( 0)& 2.8 ( 1)& 1.4 ( 2)\\
       \hline\hline
       \nucleus{\rm Fe}{56}($\overline{\nu}_e,e^+\gamma$)\nucleus{\rm Mn}{56} &
         2.3 ( 0)& 8.4 ( 0)& 1.8 ( 1)& 3.1 ( 1)& 1.4 ( 0)& 4.1 ( 0)& 1.6 ( 1)\\
       \nucleus{\rm Fe}{56}($\overline{\nu}_e,e^+$n)\nucleus{\rm Mn}{55} &
         4.2 (-1)& 4.0 ( 0)& 1.6 ( 1)& 4.0 ( 1)& 1.2 (-1)& 7.9 (-1)& 9.1 ( 0)\\
       \nucleus{\rm Fe}{56}($\overline{\nu}_e,e^+$p)\nucleus{\rm Cr}{55} &
         4.5 (-3)& 6.2 (-2)& 3.2 (-1)& 9.9 (-1)& 9.3 (-4)& 8.5 (-3)& 1.5 (-1)\\
       \nucleus{\rm Fe}{56}($\overline{\nu}_e,e^+\alpha$)\nucleus{\rm V}{52} &
         1.1 (-3)& 1.7 (-2)& 9.5 (-2)& 3.1 (-1)& 2.0 (-4)& 2.1 (-3)& 4.3 (-2)\\
       \nucleus{\rm Fe}{56}($\overline{\nu}_e,e^+$)X &
         2.8 ( 0)& 1.2 ( 1)& 3.4 ( 1)& 7.2 ( 1)& 1.6 ( 0)& 4.9 ( 0)& 2.5 ( 1)\\
    \end{tabular}
  \end{center}
  \label{tab:fe2}
\end{table}

\subsection{Lead}

The  Fermi and Ikeda sum rules both scale with neutron excess $(N-Z)$.
As the charged-current response induced by supernova $\nu_e$ neutrinos
(with average energies around 12 MeV) are dominated by Fermi and GT
transitions,
the charged-current cross sections on lead is expected to be
significantly larger than on other materials like iron or carbon.
This makes
$^{208}$Pb an attractive target for a supernova neutrino
detector.

To calculate the relevant neutrino-induced cross sections on $^{208}$Pb
we note that
convergent shell-model calculations of the GT strength
distribution are  not computationally feasible. 
Thus, unlike in $^{40}$Ar and $^{56}$Fe, 
the $\lambda^\pi =1^+$ response has been evaluated
within the RPA approach
which fulfills the Fermi and Ikeda sum rules.
As the $S_{\beta^+}$
strength (in this direction a proton is changed into a neutron)
is strongly suppressed for $^{208}$Pb, the Ikeda
sum rule fixes the $S_{\beta^-}$ strength. 
In the calculation 
the $\lambda^\pi = 1^+$ strength in $^{208}$Pb was renormalized by
the universal quenching factor which, due to a very slight
$A$-dependence is recommended to be $(0.7)^2$ in $^{208}$Pb
\cite{Martinez96}.
Thus, the Ikeda sum rule reads in this case
$S_{\beta_-} - S_{\beta_+} \approx S_{\beta_-} = 3 \cdot (0.7)^2 \cdot
(N-Z)$.
For the other multipole operators  no experimental evidence exists for
such a rescaling and we have used the RPA response.
The RPA calculation is described in details
in Ref. \cite{Kolbe01}, which also discusses the relevant
neutrino-induced cross sections for $^{208}$Pb assuming a
pion-decay-at-rest spectrum. 

The total and partial cross sections for charged current 
$(\nu_e,e^-)$ and (${\bar \nu}_e,e^+)$ reactions on
$^{208}$Pb are listed in Table \ref{tab_pb2}.
The $(\nu_e,e)$ cross section on $^{208}$Pb is
about 20 time larger than for $^{56}$Fe. This is caused by 
the ($N-Z$) and
by the strong Z-dependence of the Fermi function.
Furthermore, as  the 
IAS energy and the GT$_-$ strength is
above the neutron threshold in $^{208}$Bi at 6.9 MeV, 
most of the $(\nu_e,e)$ 
cross section leads to particle-unbound states 
and hence
decays by the neutron emission.  
Table \ref{tab_pb1}
summarizes the total and partial cross sections for neutral
current reactions on $^{208}$Pb. 

There have been other calculations of the neutrino-induced cross
sections on $^{208}$Pb.
The first study, performed
in \cite{Land}, has been critisized and improved in \cite{Fuller}.
These authors estimated the allowed transitions to the charged-current
and neutral-current cross sections empirically using data from (p,n)
scattering and from the M1 response to fix the Gamow-Teller
contributions to the cross section. 
Ref. \cite{Fuller} completed their cross section estimates by
 calculating the first-forbidden contributions on the basis
of the Goldhaber-Teller model. They calculated cross sections which are
somewhat larger than the RPA results of \cite{Kolbe01}. More
results have been reported in \cite{Jachowicz,Volpe,Suzuki} which agree
well with the RPA results. The later approach \cite{Suzuki}
is particularly
interesting as it uses the experimental GT distribution, which has
recently been determined in Osaka \cite{Krasznahorkay}, and the peaks of
the spin-dipole response to constrain the (Hartree-Fock + Tamm-Dancoff)
calculation. Furthermore, the spreading and quenching of the GT response
has been considered by coupling to 2p-2h configurations. Suzuki and
Sagawa obtain 
(3.2 $\cdot 10^{-39}$ cm$^2$) 
for the $(\nu_e,e^-)$ cross sections,
assuming a DAR neutrino spectrum, in close agreement of Kolbe's
RPA result
(3.62 $\cdot 10^{-39}$ cm$^2$) \cite{Kolbe01}.

\begin{figure}[ht]
  \begin{center}
    \leavevmode
    \includegraphics[width=0.6\columnwidth,height=12.5cm,angle=90]{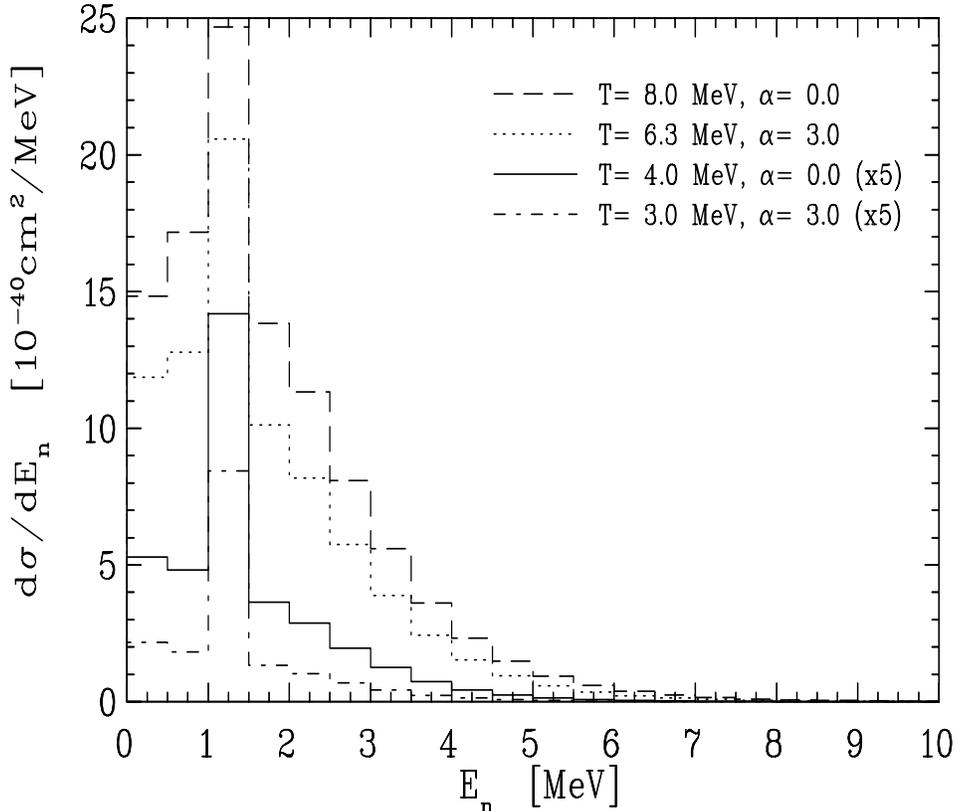}
    \caption{Neutron energy spectrum produced by the charged-current
($\nu_e,e^-)$ reaction on $^{208}$Pb. The calculation has been performed
for different supernova neutrino spectra characterized by the parameters
$(T,\alpha)$.
Note that the cross sections for $(T,\alpha)=(4,0)$ and (3,3)
have been scaled by a factor of 5.
        }
    \label{fig_pb5}    
  \end{center}
\end{figure}

Proposed detectors like LAND and OMNIS will detect the neutrons produced by the
neutrino-induced reactions on $^{208}$Pb. An obvious, but not very
sensitive  neutrino signal is the total neutron count rate. The two
detectors might also be capable of detecting
the neutron energy spectrum following the decay of states in the
daughter nucleus after excitation by charged- and neutral-current
neutrino reactions. The expected spectra have been predicted in
\cite{Kolbe01}.
The neutron spectrum for the charged current reaction on $^{208}$Pb is
dominated by the Fermi transition to the IAS and by the GT$_-$
transitions. To understand the neutron spectrum one has to consider the
neutron threshold energies for one-neutron decay (6.9 MeV) and for
two-neutron decay (14.98 MeV) in $^{208}$Bi. Hence the IAS and 
the collective
GT resonance (with an excitation energy of about 16 MeV) will decay
dominantly by 2n emission, while the low-lying GT$_-$ resonance 
at $E_x=8$ MeV decays by
the emission of one neutron. This has significant consequences for the
neutron spectrum. In the 2-neutron decay the available energy is shared
between the two emitted particles, leading 
to a rather broad and
structureless neutron energy distribution. 
As can be seen in Fig. \ref{fig_pb5} 
this broad structure is
overlaid with a peak at neutron energy around $E_n=1$ MeV caused by the
one-neutron decay of the lower GT$_-$ transition. One
expects that, due to fragmentation which is not properly described in the RPA
calculation,  the width of this peak might be broader than the 0.5
MeV-binning that has been  assumed in Fig. \ref{fig_pb5}.
The relative height of the peak compared with the broad
structure stemming from the 2n-emission is more pronounced for the
$(T,\alpha)=(4,0)$ neutrino distribution than for a potential
$(T,\alpha)=(8,0)$ $\nu_e$ spectrum as it might arise after complete
$\nu_e \leftrightarrow \nu_\mu$ oscillations.

Ideally,
OMNIS and LAND should have the ability to detect potential neutrino
oscillations. However, as has been shown in
\cite{Fuller}, the total neutron rate is by itself not suitable
to detect neutrino oscillations, even if results from various
detectors with different material 
(hence different ratios of charged-to-neutral current cross sections, 
as discussed above) are combined.
Ref. \cite{Fuller} points out that in the case of $^{208}$Pb an
attractive signal might emerge. Due to the fact that the IAS and large
portions of the GT$_-$ strength resides in $^{208}$Bi just above the
2-neutron emission threshold, Fuller {\it et al.} discuss that the
2-neutron emission rate is both flavor-specific and very sensitive to
the  temperature of the $\nu_e$ distribution.
To quantify this argument, in Ref. \cite{Kolbe01}
the cross sections for the
$^{208}$Pb($\nu_e, e^- 2n$)$^{206}$Bi reaction 
was calculated in a  model combining the
RPA for the neutrino-induced response with the statistical model for the
decay of the daughter states. The partial cross sections of
$55.7 \times 10^{-42}$ cm$^2$ 
for $\nu_e$ neutrinos with 
$(T,\alpha)=(4,0)$ Fermi-Dirac distributions should
increase significantly if neutrino
oscillations occur, as pointed out by \cite{Fuller}. 
For example, one finds for total $\nu_e
\leftrightarrow \nu_\mu$ oscillations partial 2n cross sections of
$1560 \times 10^{-42}$ cm$^2$ 
(for neutrino distributions with
parameters $(T,\alpha)= (8,0)$). We remark
that these numbers will be probably reduced, if correlations beyond the
RPA are taken into account, as part of the GT$_-$ distribution might be
shifted below the 2n-threshold.

One has to note that this 2-neutron signal will
compete with the 2-neutron decay stemming from the neutral-current reaction and
hence will reduce the flavor-sensitivity of the signal. 
Ref. \cite{Kolbe01} predicts that
the combined 2n-signal resulting from neutral-current reactions 
for the 4 $\nu_x$
neutrino types is larger than the one from the charged-current
reactions. However, if neutrino oscillations occur the neutral-current
signal is unaffected while the charged-current signal is drastically
enhanced. This supports the suggestions of Ref.
\cite{Fuller} that the 2n-signal for $^{208}$Pb detectors might be an
interesting neutrino oscillation signal.

\begin{table}[thb]
  \begin{center}
  \caption{Total cross sections for charged current
           neutrino scattering on nuclei for different neutrino energy
           spectra represented as Fermi-Dirac distributions. The
           cross sections are given in units of $10^{-42} {\rm cm}^2$.}
    \begin{tabular}{|l|c|c|c|c|c|c|c|}  
      \rule[-1ex]{0em}{4ex} { \hspace{1.8em}} ($T, \alpha$) &  
       (4,0) & (6,0) & (8,0) & (10,0) & (3,3) & (4,3) & (6.26,3) \\
       \hline \hline
       \nucleus{\rm Pb}{208}($\nu_e,e^-\gamma$)\nucleus{\rm Bi}{208} &
         4.7 ( 1)& 1.3 ( 2)& 2.5 ( 2)& 4.0 ( 2)& 3.5 ( 1)& 7.6 ( 1)& 2.2 ( 2)\\
       \nucleus{\rm Pb}{208}($\nu_e,e^-$n)\nucleus{\rm Bi}{207} &
         2.3 ( 2)& 9.9 ( 2)& 2.3 ( 3)& 4.0 ( 3)& 1.2 ( 2)& 4.2 ( 2)& 1.9 ( 3)\\
       \nucleus{\rm Pb}{208}($\nu_e,e^-$p)\nucleus{\rm Pb}{207} &
         1.8 (-2)& 1.1 (-1)& 3.3 (-1)& 6.9 (-1)& 7.2 (-3)& 3.3 (-2)& 2.3 (-1)\\
       \nucleus{\rm Pb}{208}($\nu_e,e^-\alpha$)\nucleus{\rm Tl}{204} &
         2.1 (-2)& 2.6 (-1)& 1.1 ( 0)& 3.0 ( 0)& 4.7 (-3)& 4.1 (-2)& 6.0 (-1)\\
       \nucleus{\rm Pb}{208}($\nu_e,e^-$)X &
         2.8 ( 2)& 1.1 ( 3)& 2.5 ( 3)& 4.5 ( 3)& 1.6 ( 2)& 4.9 ( 2)& 2.1 ( 3)\\
       \hline\hline
       \nucleus{\rm Pb}{208}($\overline{\nu}_e,e^+\gamma$)\nucleus{\rm Tl}{208} &
         5.8 (-1)& 3.0 ( 0)& 7.9 ( 0)& 1.5 ( 1)& 2.7 (-1)& 1.1 ( 0)& 6.1 ( 0)\\
       \nucleus{\rm Pb}{208}($\overline{\nu}_e,e^+$n)\nucleus{\rm Tl}{207} &
         4.9 (-1)& 3.8 ( 0)& 1.5 ( 1)& 3.9 ( 1)& 2.0 (-1)& 8.9 (-1)& 8.5 ( 0)\\
       \nucleus{\rm Pb}{208}($\overline{\nu}_e,e^+$p)\nucleus{\rm Hg}{207} &
         1.7 (-7)& 1.4 (-5)& 2.2 (-4)& 1.5 (-3)& 8.4 (-9)& 3.2 (-7)& 4.2 (-5)\\
       \nucleus{\rm Pb}{208}($\overline{\nu}_e,e^+\alpha$)\nucleus{\rm Au}{204} &
         4.3 (-7)& 4.0 (-5)& 6.5 (-4)& 4.4 (-3)& 2.1 (-8)& 8.1 (-7)& 1.2 (-4)\\
       \nucleus{\rm Pb}{208}($\overline{\nu}_e,e^+$)X &
         1.1 ( 0)& 6.8 ( 0)& 2.3 ( 1)& 5.4 ( 1)& 4.7 (-1)& 1.9 ( 0)& 1.5 ( 1)\\
    \end{tabular}
  \end{center}
  \label{tab_pb2}
\end{table}

\begin{table}[htb]
  \caption{Total cross sections for neutral current
                    neutrino scattering on $^{208}$Pb
for different neutrino energy
                    spectra represented as Fermi-Dirac distributions. The
                    cross sections are in units of
                    $10^{-42} {\rm cm}^2$ and are averaged over neutrinos and
                    antineutrinos.}
   \begin{center}
   \begin{tabular}{|l|c|c|c|c|c|c|c|}
      \rule[-1ex]{0em}{4ex} { \hspace{1.8em}} ($T, \alpha$) &
       (4,0) & (6,0) & (8,0) & (10,0) & (3,-3) & (4,-3) & (6.26,-3) \\
       \hline \hline
       \nucleus{\rm Pb}{208}($\nu,\nu^{\prime} \gamma$)\nucleus{\rm Pb}{208} &
         1.4 ( 0)& 7.4 ( 0)& 2.1 ( 1)& 4.5 ( 1)& 7.0 (-1)& 2.5 ( 0)& 1.5 ( 1)\\
       \nucleus{\rm Pb}{208}($\nu,\nu^{\prime}$n)\nucleus{\rm Pb}{207} &
         1.2 ( 1)& 4.8 ( 1)& 1.2 ( 2)& 2.3 ( 2)& 6.9 ( 0)& 2.0 ( 1)& 9.4 ( 1)\\
       \nucleus{\rm Pb}{208}($\nu,\nu^{\prime}$p)\nucleus{\rm Tl}{207} &
         1.6 (-5)& 3.5 (-4)& 2.4 (-3)& 8.7 (-3)& 2.9 (-6)& 3.1 (-5)& 9.0 (-4)\\
       \nucleus{\rm Pb}{208}($\nu,\nu^{\prime} \alpha$)\nucleus{\rm Hg}{204} &
         7.8 (-5)& 3.0 (-3)& 2.6 (-2)& 1.1 (-1)& 8.1 (-6)& 1.5 (-4)& 7.9 (-3)\\
       \nucleus{\rm Pb}{208}($\nu,\nu^{\prime}$)X &
         1.3 ( 1)& 5.6 ( 1)& 1.4 ( 2)& 2.7 ( 2)& 7.6 ( 0)& 2.3 ( 1)& 1.1 ( 2)\\
    \end{tabular}
  \end{center}
 \label{tab_pb1}
\end{table}

\section{Neutrino nucleosynthesis}

When the flux of neutrinos generated by the cooling of the neutron star
in a type II supernova
passes through the overlying shells of heavy elements, substantial
nuclear transmutations  are induced, despite the extremely small
neutrino-nucleus cross sections. Specific nuclei 
(e.g. $^{10,11}$B, $^{15}$N,
$^{19}$F) might be, by a large fraction, 
made by this neutrino nucleosynthesis \cite{Woosley90,Timmes95}. 
These are the product of
reaction sequences induced by neutral current $(\nu,\nu'$) reactions on
very abundant nuclei such as $^{12}$C, $^{16}$O or $^{20}$Ne. 
If the inelastic excitation of these nuclei proceeds to particle-unbound
levels, they will decay by emission of protons or neutrons, in this way
contributing to nucleosynthesis. 
As the nucleon thresholds are relatively high, effectively only
$\nu_\mu,\nu_\tau$ neutrinos and their antiparticles with their higher
average energies  contribute to the neutrino nucleosynthesis of these
elements. 
It has been noted that the neutrino induced nucleosynthesis, 
the so-called
$\nu$-process, might also be responsible for the production
of $^{138}$La and to a fraction of the $^{180}$Ta abundance
\cite{Woosley90}. The
$^{138}$La nuclide is of special interest as it appears to be 
produced by the
charged current $(\nu_e,e^-)$ reaction on
the $s$-process element $^{138}$Ba, as has already been proposed in
\cite{Woosley90,Goriely01} and recently been confirmed in detailed
studies by Heger {\it et al.} \cite{Heger03}. 
This finding is quite welcome as it makes the $\nu$ process sensitive to
the flux and distribution of supernova $\nu_e$ and $\nu_\mu,\nu_\tau$,
complementing the constraints for supernova ${\bar \nu_e}$
neutrinos from their observation in the water \v{C}erenkov detectors for
SN1987A.
Neutrino nucleosynthesis is thus potentially an important test for
the predictions of supernova models. This test is particularly
stringent, if neutrino oscillations involving $\nu_e$ neutrinos occur in
the supernova environment.

\begin{figure}[ht]
  \begin{center}
    \leavevmode
    \includegraphics[width=0.6\columnwidth,height=12.5cm]{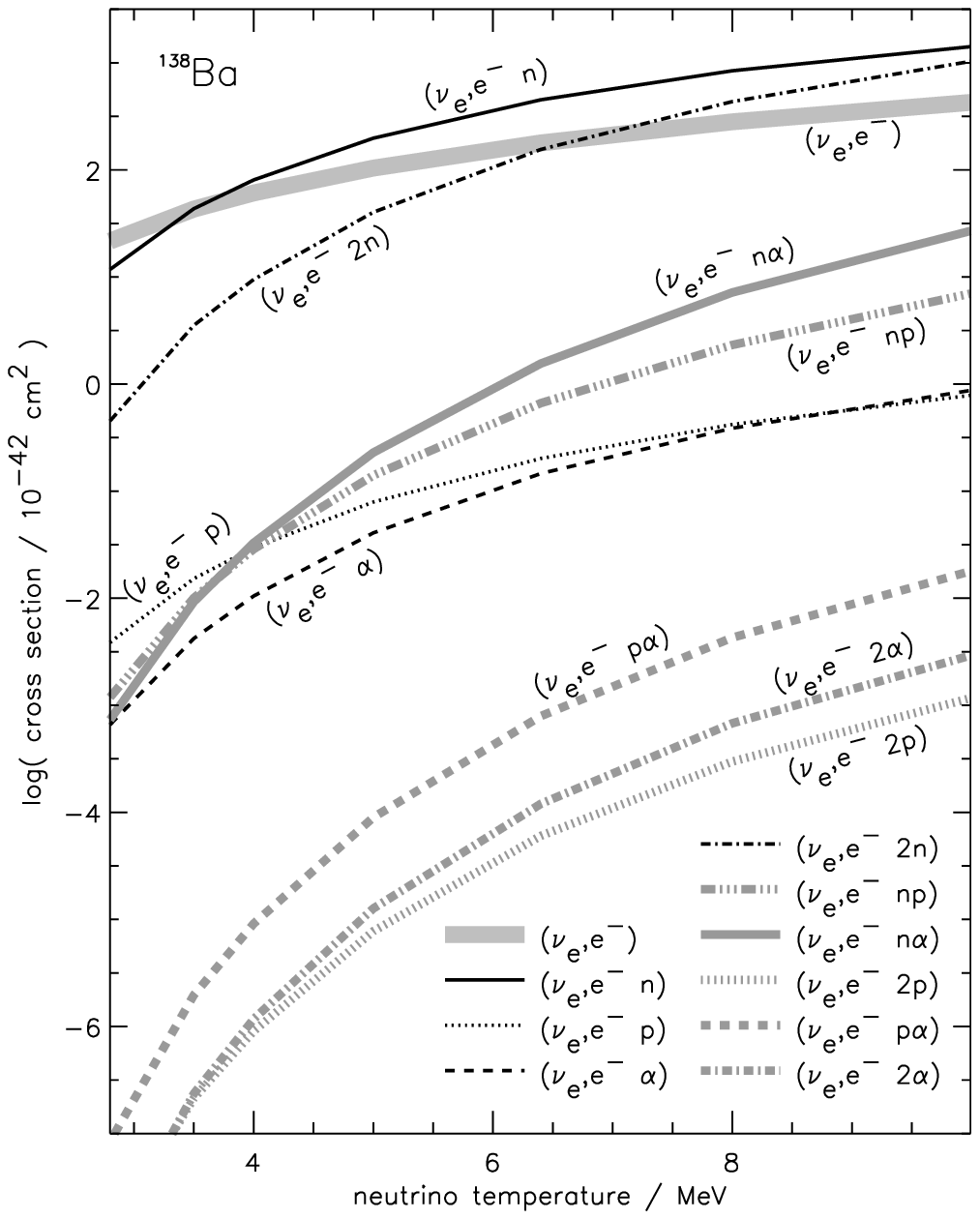}
    \caption{Partial cross sections for the $\nu_e$-induced charged-current
reaction on $^{138}$Ba. The calculations were performed for
Fermi-Dirac neutrino spectra with $\alpha=0$ and different temperature
values (from \protect\cite{Heger03}).
        }
    \label{fig_ba138}    
  \end{center}
\end{figure}

Neutrino nucleosynthesis has been proposed by Woosley {\it et al.} 
\cite{Woosley90} who
also performed detailed production studies in a 20 $M_\odot$ star. Later
Timmes {\it et al.} have extended this calculation to studies of the
$\nu$ process in a full galactical
model \cite{Timmes95}. Very recently, Heger {\it et al.} \cite{Heger03}
have improved
these earlier studies by considering mass loss in the
evolution of the progenitor stars, and by using a complete and updated reaction
network which includes all the heavy elements through bismuth. This
improvement is essential for a consistent treatment of the s-process in
the progenitor star. Last, but not least the studies of \cite{Heger03} also
used improved neutrino-induced reaction cross sections for the key
nuclei of the $\nu$ process. For the $p$ and $sd$ shell nuclei
the allowed neutrino response has been evaluated on the basis of the
shell model, while the forbidden responses were calculated within the
framework of the RPA. For the heavier nuclei, in particular for the
progenitors of $^{138}$La and $^{180}$Ta, all cross sections were
derived within the RPA. The various partial decay cross sections were
then evaluated using a statistical model cascade. 

\begin{figure}[ht]
  \begin{center}
    \leavevmode
    \includegraphics[width=0.6\columnwidth,height=10.0cm]{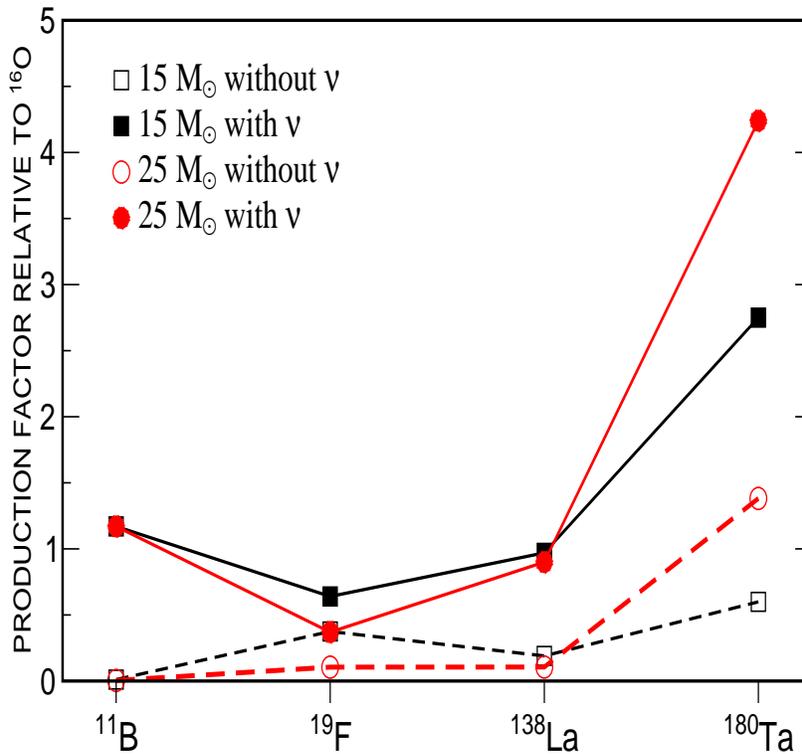}
    \caption{Production factor of $^{11}$B, $^{19}$F, $^{138}$La and
$^{180}$Ta relative to $^{16}$O in 15 $M_\odot$ (squares) and 25 $M_\odot$
(circles) stars (from \protect\cite{Heger03}). The open (filled)  symbols
represent stellar evolution studies in which neutrino reactions on
nuclei were excluded (included).
        }
    \label{fig:prod}    
  \end{center}
\end{figure}

As stressed above,  Heger {\it et al.} \cite{Heger03} 
followed the neutrino nucleosynthesis of
the light (e.g. $^{11}$B, $^{19}$F) and heavy (e.g, $^{138}$La,
$^{180}$Ta) candidate nuclei in a self-consistent way in complete
stellar evolution models that included the evolution of all isotopes up
to bismuth from the time the star ignited central hydrogen burning
through the supernova explosion. A consistent stellar modelling,
including a sufficient nuclear network, is important for two reasons.
At first, it is essential to reliably describe the production of the
progenitor nuclides (e.g. $^{12}$C, $^{20}$Ne, $^{138}$Ba...)
in the presupernova evolution. Secondly, the nuclides
produced by neutrinos from these progenitors
 can suffer severe photo-dissociations 
due to heating by the supernova shock wave which passes the region of
neutrino nucleosynthesis after the neutrinos. 
The neutrino production factors for 4 key nuclides are summarized in
Fig. \ref{fig:prod} for 15 $M_\odot$ and 25 $M_\odot$ stars. 
The results are shown relative to the $^{16}$O
production in these stars recognizing that the solar $^{16}$O abundance
is an indicator for core-collapse supernova element production.

The calculation \cite{Heger03} 
confirms earlier studies \cite{Woosley90,Timmes95} that the
neutrino nucleosynthesis makes a large fraction of the solar $^{11}$B
and $^{19}$F abundance. For both nuclides, neutral-current reactions, on
$^{12}$C and $^{20}$Ne, induced by $\nu_x$ neutrinos are the main
production source. The new calculation predicts a somewhat smaller
production of $^{19}$F than the earlier studies. 
A measurement of the GT$_0$ response on $^{20}$Ne might be quite
desirable,
including the cascade of decays of excited states,
to better constrain the $(\nu, \nu'$) cross section on $^{20}$Ne.
As predicted in \cite{Woosley90,Goriely01}, $^{138}$La is mainly produced by
charged-current reactions on $^{138}$Ba, while the $\gamma$-process
contribution is small and the neutral-current contribution from
$^{139}$La, which had been speculated to be the possible $^{138}$La
production process via $(\nu,\nu' n)$,  is insignificant.
One observes that enough $^{138}$La is being made to explain the solar
abundance, where the s-process production of the parent $^{138}$Ba in
the $s$-process prior to the supernova plays an essential role.
The calculations also show a significant production of 
the rarest nuclide $^{180}$Ta, where both, charged- and neutral-current
processes contribute. The $^{180}$Ta production factors in 
Fig. \ref{fig:prod} adds the contributions of the 
ground state and the
$9^-$ isomeric state.  As the ground state is rather short-lived ($\sim
4.5$ h), the isomeric state is the sole contributor to the solar
$^{180}$Ta abundance. Estimates \cite{Rauscher02} indicate that at
thermal freeze-out about
$30-50 \%$ of $^{180}$Ta is in the isomeric state bringing the
production factors of Fig. \ref{fig:prod} in closer agreement with the
solar abundances.

Importantly the $^{138}$La and $^{180}$Ta production is sensitive
to the neutrino distributions. For example, if the temperature
of the $\nu_e$ spectrum increases from 4 MeV to 6 MeV
(which might correspond to a neutrino oscillation scenario)
the $^{138}$La production factor increases by a factor 2 and, in the
study of \cite{Heger03}, this nuclide  would already be overproduced.
This makes the neutrino nucleosynthesis of $^{138}$La a potentially
interesting test for neutrino oscillations, in particular for the
yet unmeasured neutrino mixing angle $\theta_{13}$ \cite{Heger03}.

Finally, we note that 
$\nu$ nucleosynthesis of $^{138}$La and $^{180}$Ta competes with
the p-process production of these elements, making a reliable determination
of the p-process abundances also important.

\section{Conclusions}

In this review we have described the methods used in the evaluation
of the neutrino-nucleus reaction cross sections. We have shown,
in detail, the
results of calculations of these cross sections for a variety of targets
of practical importance. 

Many of the described reactions are accessible in experiments with 
a neutrino source  from the pion and muon decay at rest,
available at the future very intense 
neutron spallation sources. Detailed comparison
of the results of such experiments would establish important benchmarks
for comparison of theory and experiment. This, in turn, would lead
to refinements or possible modifications of the theoretical treatment
of processes involving the charged and neutral current interaction
of neutrinos with complex nuclei. Having a reliable tool 
for such calculation is of great importance in a variety of
applications, like the study of neutrino oscillations, detection
of supernova neutrino signal, description of the neutrino transport
in supernovae, or description of the $r$-process nucleosynthesis.

\vspace{1cm}

{\bf Acknowledgment}

The work of PV was supported in part by the US DOE and by the 
Institute of Physics and Astronomy at the University of {\AA}rhus.
KL is partially supported by the Danish Research Foundation.
GMP is supported by the Spanish MCyT under
contracts AYA2002-04094-C03-02 and AYA2003-06128.


\begin{thebibliography}{\hspace{1.5cm}}


\bibitem{Walecka} J.D. Walecka, Semi-leptonic weak interactions in
nuclei, in {\it Muon physics}, eds. V.W. Hughes and C.S. Wu (Academic Press,
New York, 1972) 

\bibitem{Co72} J. S. O'Connell, T. W. Donnelly, and J. D. Walecka,
Phys. Rev. C{\bf 6}, 719 (1972).


\bibitem{Don-Hax} T. W. Donnelly and W. C. Haxton, At. Data and Nucl. Data.
Tables, {\bf 23}, 103 (1979).

\bibitem{DP79} T. W. Donnelly and R. D. Peccei, Phys. Rep. {\bf 50}, 1 (1979).

\bibitem{Kolbe96} E. Kolbe, Phys. Rev. C{\bf 54}, 1741 (1996).

\bibitem{Be82} H. Behrens and W. B\"{u}hring, {\it Electron Radial
Wave Functions and Nuclear Beta-Decay}, (Clarendon, Oxford, 1982).  

\bibitem{Engel98} J. Engel, Phys Rev. C{\bf 57}, 2004 (1998).

\bibitem{Caurier04} E. Caurier, G. Martinez-Pinedo, F. Nowacki, A. Poves
and A. Zuker, Rev. Mod. Phys., in print
 
\bibitem{CRPA} M. Buballa, S. Dro\.{z}d\.{z}, S. Krewald, and
J. Speth, Ann. Phys. {\bf 208}, 346 (1991).

\bibitem{Kolbe92} E. Kolbe, K. Langanke, S. Krewald and F.-K. Thielemann,
Nucl. Phys. {\bf A540}, 599 (1992).

\bibitem{Moniz} R. A. Smith and E. J. Moniz, Nucl. Phys. {\bf B43},
605 (1972).

\bibitem{Friedel} J.J. Cowan, F.K. Thielemann and J.W. Truran, Phys.
Rep. 208 (1991) 208; T. Rauscher and F.-K. Thielemann, At. Data Nucl.
Data Tables {\bf 75}, 1 (2000); {\bf 79}, 47 (2001.


\bibitem{Langanke96} K. Langanke, P. Vogel and E. Kolbe, Phys. Rev. Lett.
{\bf 76}, 2629 (1996).


\bibitem{Haxton87} W. C. Haxton, Phys. Rev. D{\bf 36}, 2283 (1987).

\bibitem{Kolbe94} E. Kolbe, K. Langanke and P. Vogel, Phys. Rev. {\bf
C50}, 2576 (1994).

\bibitem{footnote} The partial rates listed in Table II of Ref.
\cite{Kolbe94} should be multiplied by a factor 1000.

\bibitem{Karmen}  G. Drexlin et al., KARMEN collaboration,
               Phys. Lett. {\bf B267}, 321 (1991);
B. Zeitnitz, KARMEN Collaboration,
               Prog. Part. Nucl. Phys. {\bf 32}, 351 (1994);
J. Kleinfeler et al. KARMEN collaboration, in
{\em Neutrino'96}, eds. K. Enquist, K. Huitu, and J. Maalampi,
{World Scientific, Singapore, 1997}. 
\bibitem{LSND}  M. Albert et al., Phys. Rev. {\bf C51}, 1065 (1995);
C. Athanassopoulos et al., Phys. Rev. C {\bf 55}, 2078 (1997); 
C. Athanassopoulos et al., Phys. Rev. C {\bf 56}, 2806 (1997); 
R. Imlay, Nucl. Phys. {\bf A629}, 531c (1998).
\bibitem{Krakauer} D.A. Krakauer et al., Phys. Rev. {\bf C45}, 2450 (1992).
\bibitem{Kolbe94} E. Kolbe, K. Langanke, and S. Krewald,
               Phys. Rev. {\bf C49}, 1122 (1994).
\bibitem{Kolbe95}E. Kolbe, K. Langanke, F.-K. Thielemann, and P. Vogel, 
               Phys. Rev. {\bf C52}, 3437 (1995).
\bibitem{Kolbe99a} E. Kolbe, K. Langanke and P. Vogel, Nucl. Phys. {\bf
A652}, 91 (1999)

\bibitem{Oset} T. S. Kosmas and E. Oset, Phys. Rev.C {\bf 53}, 1409 (1996).
\bibitem{Oset98} S. K. Singh, N. C. Mukhopadhyay, and E. Oset,
Phys. Rev. C {\bf 57}, 2687 (1998).
\bibitem{Mintz} S. L. Mintz and M. Pourkaviani, 
Nucl. Phys. {\bf A594}, 346 (1995).
\bibitem{Auerbach} N. Auerbach, N. Van Giai, and O. K. Vorov, 
Phys. Rev. C {\bf 56}, R2368 (1997).
\bibitem{triad} J. Engel, E. Kolbe, K. Langanke, and P. Vogel,
                        Phys. Rev. {\bf C54}, 2740 (1996).
\bibitem{Fukugita} M. Fukugita, Y. Kohyama and K. Kubodera,
                 Phys. Lett. {\bf B212}, 139 (1988).

\bibitem{Leiss}  J. E. Leiss and R. E. Taylor, 
as quoted in W. Czyz, Phys. Rev. {\bf 131}, 2141 (1963).

\bibitem{Kolbe96p} E. Kolbe, K. Langanke and P. Vogel, Nucl. Phys. {\bf
A613}, 382 (1996)



\bibitem{Hayes01} A. C. Hayes and I. S. Towner, Phys. Rev. C {\bf 61},
0044604 (2001).

\bibitem{Brown02} N. Auerbach and B. A. Brown, Phys. Rev. C {\bf 65},
024322 (2002).

\bibitem{Raffelt} G. Raffelt {\it Stars as Laboratories for Fundamental
Physics} (Chicago Press, 1996)

\bibitem{Hirata} K.S. Hirata {\it et al.}, Phys. Rev. Lett. {\bf 58}, 1490
(1987).

\bibitem{IMB} R.M. Bionta {\it et al.}, Phys. Rev. Lett. {\bf 58}, 1494 (1987).

\bibitem{Omnis} D.B. Cline {\it et al.}, Astrophys. Letters and
Communications {\bf 27},403 (1990); Phys. Rev. {\bf D50}, 720 (1994). 

\bibitem{Land} C.K. Hargrove {\it et al.}, Astroparticle Physics {\bf
5}, 183 (1996).

\bibitem{SNO} G. Ewan, NIM {\bf A314}, 373 (1992); SNO Collaboration
Physics in Canada {\bf 48}, 112 (1992).

\bibitem{Keil02} M. Th. Keil, G. Raffelt and H-T. Janka, astro-ph/0208035.

\bibitem{Janka} H.-T. Janka and W. Hillebrandt, Astron. Astrophys. {\bf
224},49 (1989); Astron. Astrophys. Suppl. {\bf 78}, 375 (1989).

\bibitem{Wilson} J.R. Wilson (private communication), as cited in
Y.-Z. Qian, W.C. Haxton, K. Langanke and P. Vogel, Phys. Rev. {\bf C55},
1532 (1997).  

\bibitem{Kolbe99} E. Kolbe, K. Langanke and G. Martinez-Pinedo, Phys. Rev.
{\bf C60}, 052801 (1999).

\bibitem{Hektor} A. Hektor, E. Kolbe, K. Langanke, and J. Toivanen,
Phys. Rev. {\bf C61}, 055803 (2000).

\bibitem{Totsuka} Y. Totsuka, Rep. Progr. Phys. {\bf 55}, 377 (1992) .

\bibitem{Nussinov} S. Nussinov and R. Shrock, Phys. Rev. Lett. {\bf 86},
2223 (2001).

\bibitem{Kolbe02} E. Kolbe, K. Langanke and P. Vogel, Phys. Rev.
{\bf D66}, 013007 (2002).

\bibitem{Engel} J. Engel, E. Kolbe, K. Langanke and P. Vogel, Phys. Rev.
{\bf C48}, 3048 (1993).

\bibitem{Icarus} F. Cavanna and P. Palamara, Nucl. Phys. {\bf B112}, 265
(2002) 

\bibitem{Ormand} W.E. Ormand, Pizzechero, P.F. Bortignon and R.A.
Broglia, Phys. Lett {\bf B345}, 343 (1995) 


\bibitem{Maschuw} R. Maschuw, Prog. Part. Phys. {\bf 40}, 183 (1998); K.
Eitel, private communication.

\bibitem{Toivanen} J. Toivanen, E. Kolbe, K. Langanke, G. Martinez-Pinedo
and P. Vogel, Nucl. Phys. {\bf A694}, 395 (2001)

\bibitem{Haxton88} W.C. Haxton, Phys. Rev. Lett. {\bf 60}, 1999 (1988)

\bibitem{rmp} K. Langanke and G. Martinez-Pinedo, Rev. Mod. Phys. {\bf
75}, 819 (2003)

\bibitem{Langanke01a} K. Langanke, G. Martinez-Pinedo and J.M. Sampaio,
Phys. Rev. {\bf C64}, 055801 (2001)

\bibitem{Sampaio02} J.M. Sampaio, K. Langanke, G. Martinez-Pinedo and
D.J. Dean, Phys. Lett. {\bf B529}, 19 (2002)

\bibitem{Martinez96} G. Martinez-Pinedo, A. Poves, E. Caurier and A.P.
Zuker, Phys. Rev. {\bf C53}, R2602 (1996)

\bibitem{Kolbe01} E. Kolbe and K. Langanke, Phys. Rev. C {\bf 63},
025802 (2001).

\bibitem{Fuller} G.M. Fuller, W.C. Haxton and G.C. McLaughlin,
Phys. Rev. {\bf D59}, 085005 (1999).

\bibitem{Jachowicz} N. Jachowitz, K. Heyde, and J. Ryckebush,
Phys. Rev. C {\bf 66}, 055501 (2000).

\bibitem{Volpe} C. Volpe, N. Aurbach, G. Colo, and N. Van Giai,
 Phys. Rev. {\bf C65}, 044603 (2002); J. Engel, G. C. McLaughlin,
and C. Volpe,  Phys. Rev. {\bf C67}, 013005 (2003). 

\bibitem{Suzuki} T. Suzuki and H. Sagawa, Nucl. Phys. {\bf A718},
446c (2003).

\bibitem{Krasznahorkay} A.Krasznahorkay {\it et al.},   
 Phys. Rev. {\bf C64}, 067302 (2001). 

\bibitem{Woosley90} S.E. Woosley, D.H. Hartmann, R.D. Hoffman and W.C.
Haxton, Astr. J. {\bf 356}, 272 (1990) 

\bibitem{Timmes95} F.X. Timmes, S.E. Woosley and T.A. Weaver, ApJS {\bf
98}, 617 (1995)

\bibitem{Goriely01} S. Goriely, M. Arnould, I. Borsov and M. Rayet,
Astr. Astrophys. {\bf 375}, 35 (2001)

\bibitem{Rauscher02} T. Rauscher, A. Heger, R.D. Hoffman and S.E.
Woosley, ApJ {\bf 576}, 323 (2002)

\bibitem{Heger03} A. Heger, E. Kolbe, W.C. Haxton, K. Langanke, G.
Martinez-Pinedo and S.E. Woosley, submitted to PRL



\end{thebibliography}
\end{document}